\documentclass[onecolumn, aps, floatfix, altaffilletter, superscriptaddress, preprintnumbers, showpacs, nofootinbib, showkeys, 12pt]{revtex4-1}
\pdfoutput=1
\usepackage[utf8]{inputenc}
\usepackage[T1]{fontenc}
\usepackage[colorlinks=true,citecolor=blue,linkcolor=blue]{hyperref}
\usepackage[normalem]{ulem}
\usepackage{amsmath,amssymb}
\usepackage{epsfig}
\usepackage{adjustbox}
\usepackage{graphicx}               % Standard graphics package
\usepackage{url}
\usepackage{color}
\usepackage{slashed}
\usepackage{multirow}
\usepackage{placeins}
\usepackage[dvipsnames]{xcolor}
\usepackage{epstopdf}
\usepackage{soul}
\usepackage{tikz}
\usepackage{mathtools}
\usepackage{xcolor}
\usepackage{todonotes}

\newcommand{\beq}{\begin{equation}}
\newcommand{\eeq}{\end{equation}} 
\newcommand{\bea}{\begin{eqnarray}}
\newcommand{\eea}{\end{eqnarray}}
\newcommand{\ba}{\begin{array}}
\newcommand{\ea}{\end{array}}

\definecolor{pink}{RGB}{255,105,180}

\newcommand{\ifb}{{\,{\rm fb}^{-1}}}

\begin{document}
%========================================

\title{
Seeking for sterile neutrinos with displaced leptons at the LHC}

\author{Jia Liu}
\affiliation{Enrico Fermi Institute, University of Chicago, Chicago, IL 60637, USA}

\author{Zhen Liu}
\affiliation{Maryland Center for Fundamental Physics, Department of Physics, University of Maryland, College
Park, MD 20742, USA}

\author{Lian-Tao Wang}
\affiliation{Enrico Fermi Institute, University of Chicago, Chicago, IL 60637, USA}
\affiliation{Kavli Institute for Cosmological Physics, University of Chicago, Chicago, IL 60637, USA}

\author{Xiao-Ping Wang}
\affiliation{High Energy Physics Division, Argonne National Laboratory, Argonne, IL 60439, USA}

\date{\today}
%\keywords{Supersymmetry, Beyond the Standard Model, Large Hadron Collider, precision timing}
%\pacs{95.35.+d, 14.80.Da, 14.80.Ec}

\preprint{
%     \begin{flushright}
   EFI-19-6 
 %    \end{flushright}
}
%========================================

\begin{abstract}
We study the signal of long-lived sterile neutrino at the LHC produced through the decay of  the $W$ boson.  It decays into charged lepton and jets.
The characteristic signature is a hard prompt lepton and a lepton from the displaced decay of the sterile neutrino, 
which leads to a bundle of displaced tracks with large transverse impact parameter.
Different from other
studies, we neither reconstruct the displaced vertex nor place requirement on its invariant mass
to maintain sensitivity for low sterile neutrino masses. Instead, we focus on the displaced track from the lepton. A difficulty for low mass sterile neutrino study
is that the displaced lepton is usually \textit{non-isolated}. Therefore, leptons from heavy flavor 
quark is the major source of background. We closely follow a search for 
displaced electron plus muon search at CMS and study their control regions, which is related to our signal regions, in great detail to develop a robust estimation of the background for our signals.
After further optimization on the signal limiting the number of jets, low $H_T$ and
large lepton displacement $d_0$ to suppress SM background, we reach an exclusion sensitivity of about $10^{-8}$
($10^{-5}$) for the mixing angle square at 10 (2) GeV sterile neutrino mass respectively. The strategy we propose can 
cover the light sterile masses complimentary to beam dump and forward detector 
experiments. 
\end{abstract}

\maketitle
\tableofcontents

\setcounter{secnumdepth}{2} 
\setcounter{tocdepth}{2}

%========================================

\section{Introduction}
The origin of the neutrino masses is a puzzle of the Standard Model~\cite{Fukuda:1998mi,Fukuda:1998ah,Ahmad:2002jz,Eguchi:2002dm}.
The parameters in the neutrino sector, such as the mass differences and mixing angles, have been measured with higher and higher precision~\cite{Tanabashi:2018oca}. At the same time, cosmological observations set an upper bound on the sum of neutrino masses to be smaller than $0.12$ eV and the effective extra relativistic degrees of freedom to be $N_{\rm eff} = 2.99 \pm 0.17$ from Planck 2018 data~\cite{Aghanim:2018eyx}. 
Many mechanisms have been introduced to incorporate the presence of the tiny neutrino masses. Among them, the seesaw type of solutions to the small neutrino mass is probably the most plausible. A heavier, often sterile, neutrino to realize the seesaw mechanism of a certain type would be a smoking gun signal for this class of models.

Many active experimental programs have been developed to search for sterile neutrinos, through oscillation via light 
(eV-keV) ones at current or future short and long-baseline neutrino facilities \cite{Kopp:2013vaa, Giunti:2015wnd, Capozzi:2016vac, Gariazzo:2017fdh, Dentler:2017tkw, Dentler:2018sju}, neutrinoless double-beta decay (MeV) for intermediate ones~\cite{Benes:2005hn, Atre:2009rg, Blennow:2010th, Rodejohann:2011mu, Barea:2015zfa}
and at LHC through same-sign dileptons for heavy ones (100 GeV or above) \cite{Keung:1983uu,Han:2006ip,delAguila:2007qnc, Atre:2009rg,Alva:2014gxa,Degrande:2016aje,Cai:2017mow,Accomando:2016rpc, Accomando:2017qcs,Cvetic:2018elt}. 
Furthermore, in the mass range of GeV to 100 GeV, the sterile neutrino would be metastable or long-lived at the detector scale, and can be probed at beam-dump types of experiments \cite{Badier:1985wg, Bergsma:1985is, CooperSarkar:1985nh, Bernardi:1987ek, Baranov:1992vq, Vilain:1994vg, Gallas:1994xp, Vaitaitis:1999wq, Astier:2001ck, Orloff:2002de, Adams:2013qkq, Anelli:2015pba, Drewes:2018gkc} 
(see also a review \cite{Deppisch:2015qwa}),
Very recently, the searches for sterile neutrino at the LHC started to develop actively as part of the  long-lived particles searches covering the GeV scale~\cite{Ahmad:2002jz, Graesser:2007pc, Graesser:2007yj, Helo:2013esa, Maiezza:2015lza, Batell:2016zod, Antusch:2016vyf, Nemevsek:2016enw, Antusch:2017hhu, Caputo:2017pit, Cottin:2018kmq, Helo:2018qej, Kling:2018wct, Curtin:2018mvb,Abada:2018sfh, Dib:2019ztn, Drewes:2019fou, Bondarenko:2019tss}.

In this work, 
we focus on sterile neutrino production from $W$ boson decay, which leads to a signature 
of a prompt lepton and a displaced but non-isolated lepton. A central challenge for new phenomenological studies on long-lived particles at the LHC is how to estimate the corresponding background. We overcome, at least partially, this challenge by extracting background behavior information from two very similar control regions measured and validated by an experimental search targeting different signatures at CMS~\cite{CMS-PAS-EXO-16-022}.
We show the LHC sensitivity would be improved significantly in the regime of sterile neutrino mass around 1-20 GeV with a mixing angle squared between $10^{-8}$ to $10^{-3}$.

The paper is organized as follows. In Sec.\ref{sec:theory}, we provide a brief review of the seesaw models. In Sec.\ref{sec:properties}, we discuss the properties of the sterile neutrino relevant for this study. In Sec.~\ref{sec:analysis}, we present the proposed analysis with the corresponding background estimation and the resulting for model parameter coverage. 
Finally, we conclude and discuss future directions in Sec.~\ref{sec:conclusion}.

\section{Sterile neutrino models}
\label{sec:theory}

In this section, we briefly review a few classes of seesaw models to motivate the parameter space we focus on. We begin with the original seesaw models \cite{Minkowski:1977sc, GellMann:1980vs, Mohapatra:1979ia, Yanagida:1980xy, Schechter:1980gr} with the interaction Lagrangian of the new sterile neutrino sector
\begin{align}
\Delta \mathcal{L}_{\nu} = - \lambda_\nu \bar{L} \tilde{H} N - \frac{m_N}{2} \bar{N}^c N + h.c.  \,,
\end{align}
where $\tilde{H} = i \sigma_2 H^*$. The mass matrix in the flavor basis $\left\lbrace \nu_L, N^c  \right\rbrace$ is 
\begin{align}
M_\nu = 
\left(\begin{array}{cc}
0 & m_D \\
m_D  & m_N \end{array}
\right) ,
\end{align}
where $m_D = \lambda_\nu v/\sqrt{2} $ with the vacuum expectation value of the Higgs field $v=246~$ GeV. The mass of light and heavy neutrino are
\begin{align} 
m_\nu \equiv m_1  \simeq  \frac{m_D^2}{m_N} , \quad
m_2  \simeq m_N + \frac{m_D^2}{m_N} \simeq m_N
\end{align}
respectively, in the heavy Majorana mass limit. The mixing angle is $\sin\theta = m_D/ m_N$, which yields 
a relation between the mixing angle, light  (active) neutrino mass and heavy neutrino mass:
\beq
\sin^2\theta \simeq \frac {m_\nu} {m_N}.
\label{eq:angle}
\eeq

As we shall see later, the mixing angle controls the phenomenology of the searches for the sterile neutrino $N$. Beyond the original (and the most basic) seesaw model, a number of extensions and variations have been proposed, with new parameters and deviations beyond the simple relation shown in Eq.~\ref{eq:angle}. We mention two such examples, namely, inverse seesaw model \cite{Mohapatra:1986aw, Mohapatra:1986bd} or linear seesaw model \cite{Wyler:1982dd, Akhmedov:1995ip, Akhmedov:1995vm}. These variations of seesaw models can be parameterized similarly, and they are  more accessible (and testable) with a larger mixing angle while capable of reproducing the light neutrino mass. The phenomenology is controlled by the same set of phenomenologically relevant parameters $\sin\theta$ and $m_N$ but with a different relation to Lagrangian parameters. In inverse seesaw, we have
\beq
\sin^2\theta = \frac {m_\nu} {\mu} \quad \text{(inverse seesaw)} ,
\eeq
where $\mu$ is a free parameter with the relation $m_\nu = \mu \left(\frac {m_D} {m_N}\right)^2$. Here $m_N$ becomes a Dirac mass term of new sterile neutrinos and $\mu$ is the Majorana mass of one of the new species.  For linear seesaw, the relation is
\beq
\quad \sin\theta = \frac{m_\nu}{m_\psi} \quad \text{(linear seesaw)},
\eeq
where $m_\psi$ is an additional Dirac mass term small than $m_N$, and its presence will violate the lepton number.
With the additional parameters introduced, we can attribute neutrino mass to these new parameters.
Thereafter, we only focus on $\sin\theta$ and $m_N$ which are both free parameters.

In the later sections, we will constrain the mixing angle $\sin\theta$ as a function of sterile neutrino mass $m_N$ with experimental studies. To simplify this discussion and focus on demonstrating our strategy, we will make the simple assumption that all three generations of the active neutrinos mix with $N$ equally. In the context of more general seesaw models, making this assumption does not commit us to a particular neutrino mass hierarchy. The collider signal studied in this paper mainly rely on final states with electrons and muons, while the mixing with $\tau$ neutrino enters the decay width of the sterile neutrino if $m_N > m_\tau$. If we relax the assumption of universal mixing, the reach can be obtained from our result by rescaling relevant rates. 

In the basic seesaw model, the mixing angle is not a free parameter for a particular sterile neutrino mass. It is fixed to be around 
\beq
\sin^2 \theta \simeq 10^{-12} \left(\frac{m_\nu}{0.01\ {\rm eV}} \right) \left( \frac{10 \ {\rm GeV}}{m_N}\right), 
\eeq
which is very difficult to be probed at the colliders. Still, in the context of more extended models, larger mixing angle is allowed. It is these non-minimal models which would be the main target for LHC searches.

\section{Sterile neutrino properties}
\label{sec:properties}

In this section, we discuss the features and properties of sterile neutrino from the extended seesaw
models, with mixing angle $\sin\theta$ and mass $m_N$ as free parameters.
Due to the mixing between SM neutrino $\nu$ and sterile neutrino $N$, the relevant interactions are 
\begin{align}
\mathcal{L}=\frac{g \sin\theta}{\sqrt{2}} \left( W_\mu \bar{\ell}_L \gamma^\mu  N + h.c. \right) -\frac{g\cos\theta \sin\theta}{2\cos\theta_w} Z_\mu  \left( \bar{\nu}_L \gamma^\mu  N + \bar{N} \gamma^\mu  \nu_L \right)+ \frac{g \sin^2\theta}{2\cos\theta_w}  Z_\mu \bar{N}\gamma^\mu P_L N ,
\end{align}
where $\theta_w$ is the weak mixing angle.

\begin{figure}[htb]
    \centering
    \includegraphics[width=0.32 \textwidth]{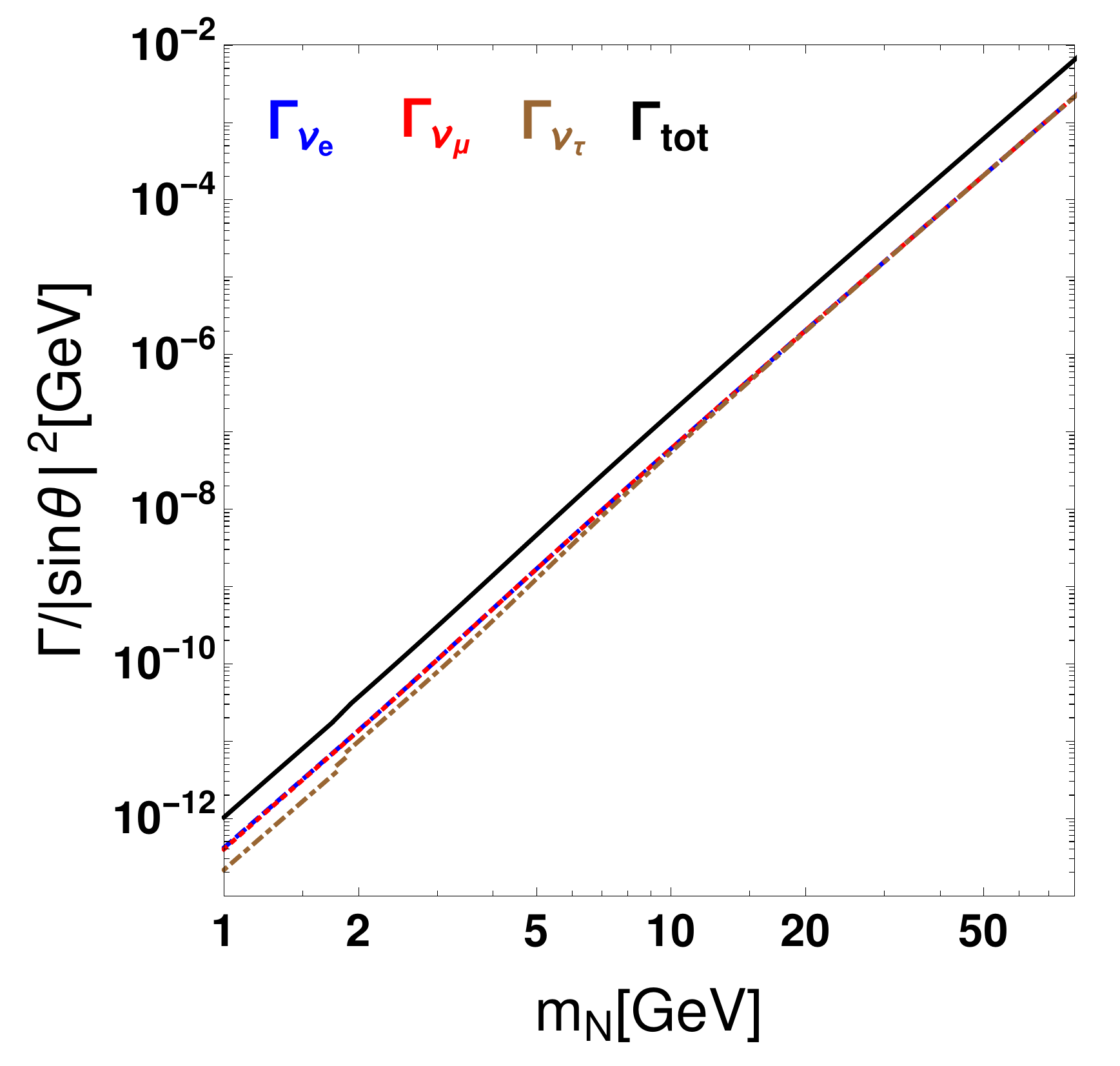}
    \includegraphics[width=0.32 \textwidth]{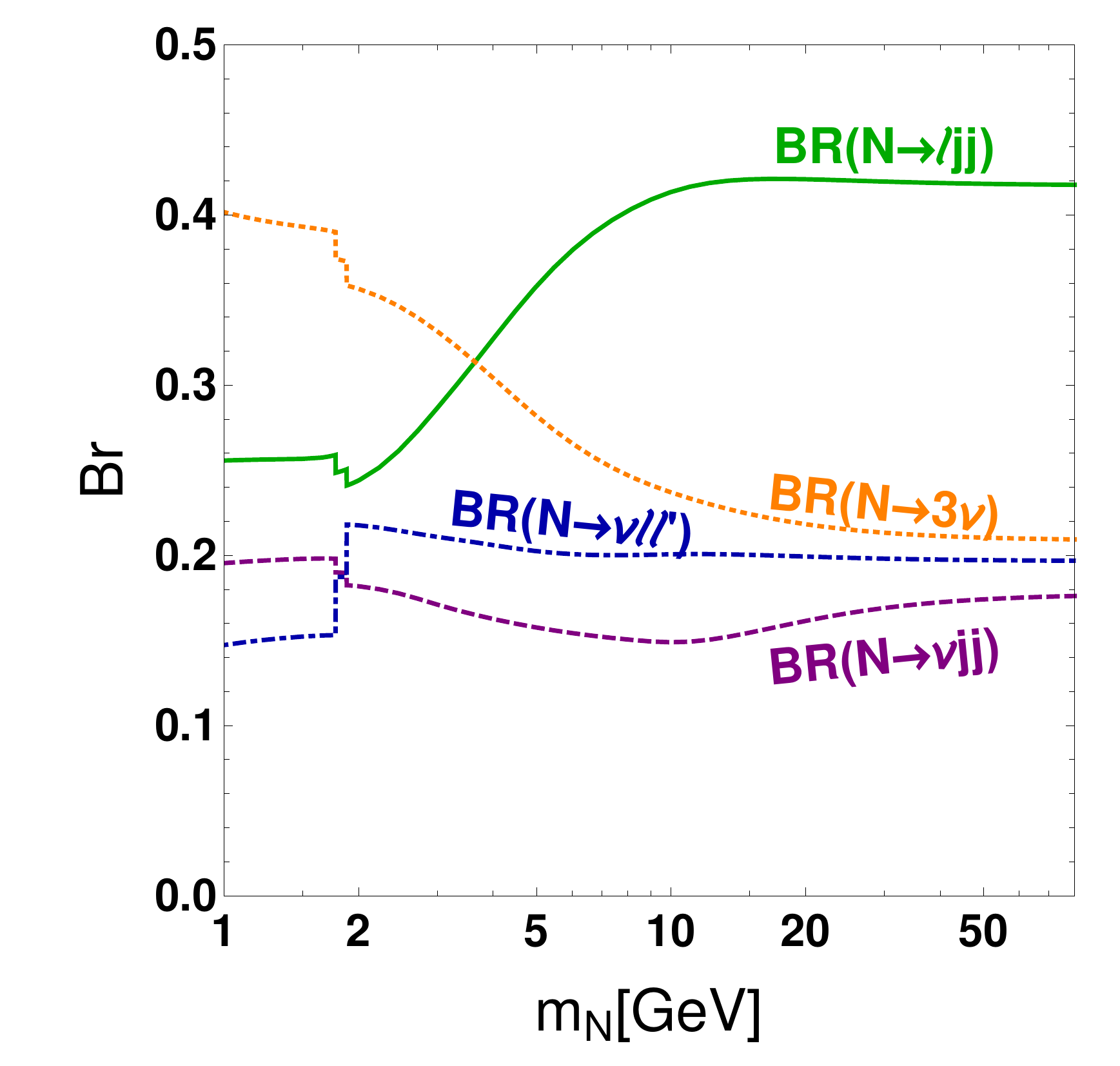}
    \includegraphics[width=0.32 \textwidth]{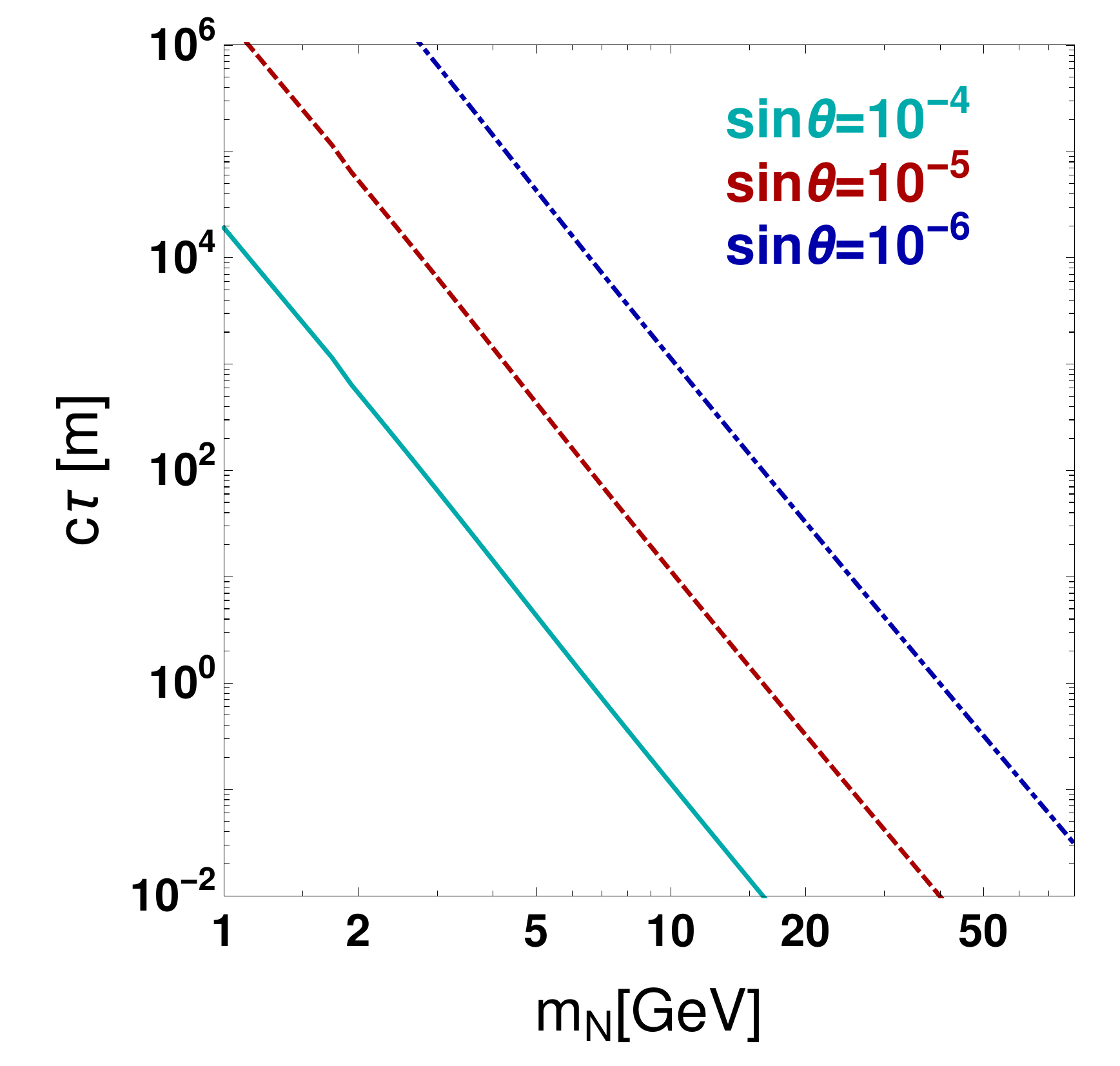}
    \caption{The decay partial widths, branching ratios and lifetime for heavy sterile neutrino $N$, under a democratic mixing assumption $\sin\theta_{\nu_e N} = \sin\theta_{\nu_\nu N} =\sin\theta_{\nu_\tau N}=\sin\theta$. In the left panel,
    we show the partial decay width from mixing with $\nu_e$, $\nu_\mu$, $\nu_\tau$ and total
    decay width respectively. In the middle panel, we show the branching ratios to leptonic channels
    $\nu \ell \ell'$, $\nu \nu \nu $ and semi-leptonic channels $\ell j j$ and $\nu j j$. In the right panel,
    we show the lifetime as a function of $m_N$ with benchmark mixing angle $\sin \theta = 10^{-4}$, $ 10^{-5}$
    and $10^{-6}$ respectively.
    }
    \label{fig:decays}
\end{figure}

In a large region of parameter spaces of interests, the right-handed neutrino has a macroscopic lifetime
\begin{equation}
    \label{eq:lifetime}
    c\tau \simeq 12 {\rm ~km} \times \left(\frac {10^{-12}} {\sin^2\theta}\right) \left(\frac {10~\rm GeV} {m_N}\right)^5 ,
\end{equation}
with the details of partial decay widths following \cite{Helo:2010cw, Helo:2013esa}, see also \cite{Bondarenko:2018ptm, Das:2018tbd}. 
The reference mixing angle squared $10^{-12}$ is chosen for SM neutrino mass to  $0.01$~eV in the basic seesaw model. For an intermediate value of the mixing angle squared $10^{-6}$, a 10 GeV sterile neutrino will still be long-lived at collider level with a proper lifetime of 12~mm.
In Fig.~\ref{fig:decays}, we show the partial widths, branching ratios and lifetime for heavy sterile neutrino $N$ decay, 
with the assumption mixing angles between the three SM flavor neutrino are democratic $\sin\theta_{\nu_e N} = \sin\theta_{\nu_\nu N} =\sin\theta_{\nu_\tau N}$.

At colliders, the production of the sterile neutrino is the same as SM neutrinos, with additional suppression from mixing angle and kinematics. The production through the on-shell $W$ boson is of particular interest here, given 
that the associated charge leptons can be used to trigger the signal processes. The long-lived 
sterile neutrino can then be analyzed without trigger penalties. The expected total number of sterile neutrino at the HL-LHC for one generation of democratic sterile neutrino is approximately
\beq
\mathcal{L}\times\sigma(pp\to W^\pm){\rm Br}(W^\pm\to\ell^\pm N) \simeq 1.8 \times 10^{5}
\left( \frac{\sin^2\theta}{10^{-6}} \right),
\eeq
where $\mathcal{L} = 3000 {\rm ~fb}^{-1}$ is the integrated luminosity at the HL-LHC.

\begin{figure}[htb]
    \centering
    \includegraphics[width=0.45 \textwidth]{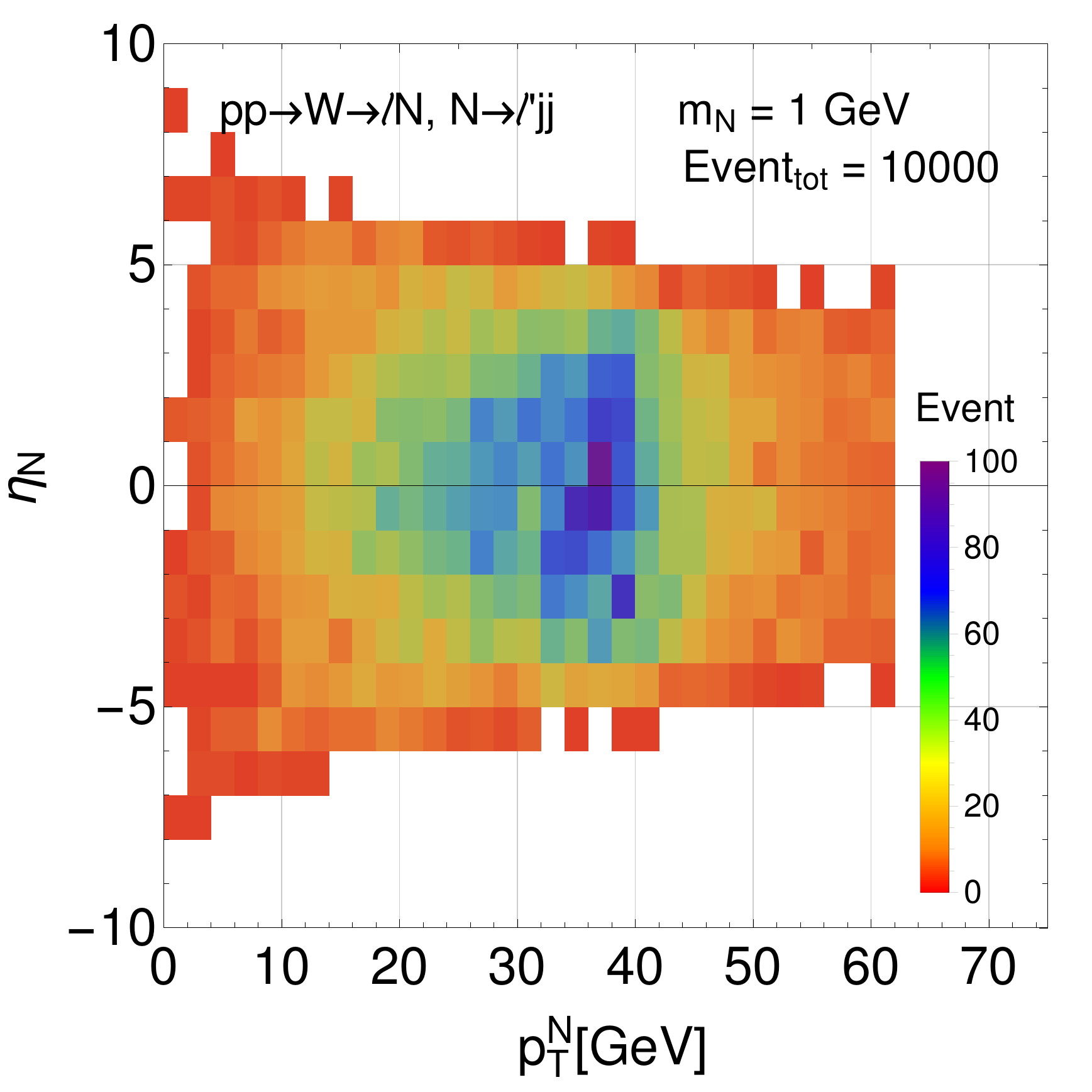}
    \includegraphics[width=0.45 \textwidth]{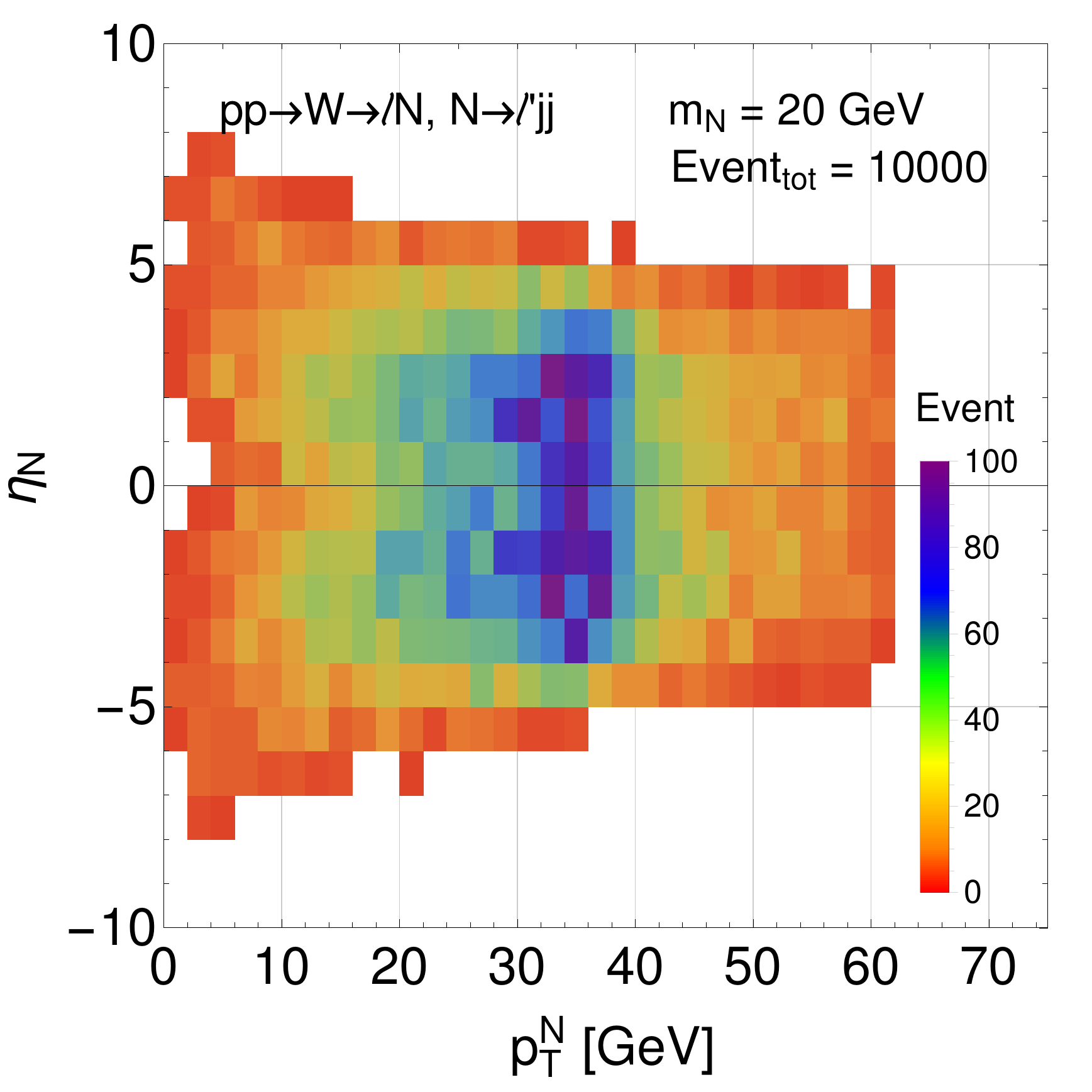}
    \includegraphics[width=0.45 \textwidth]{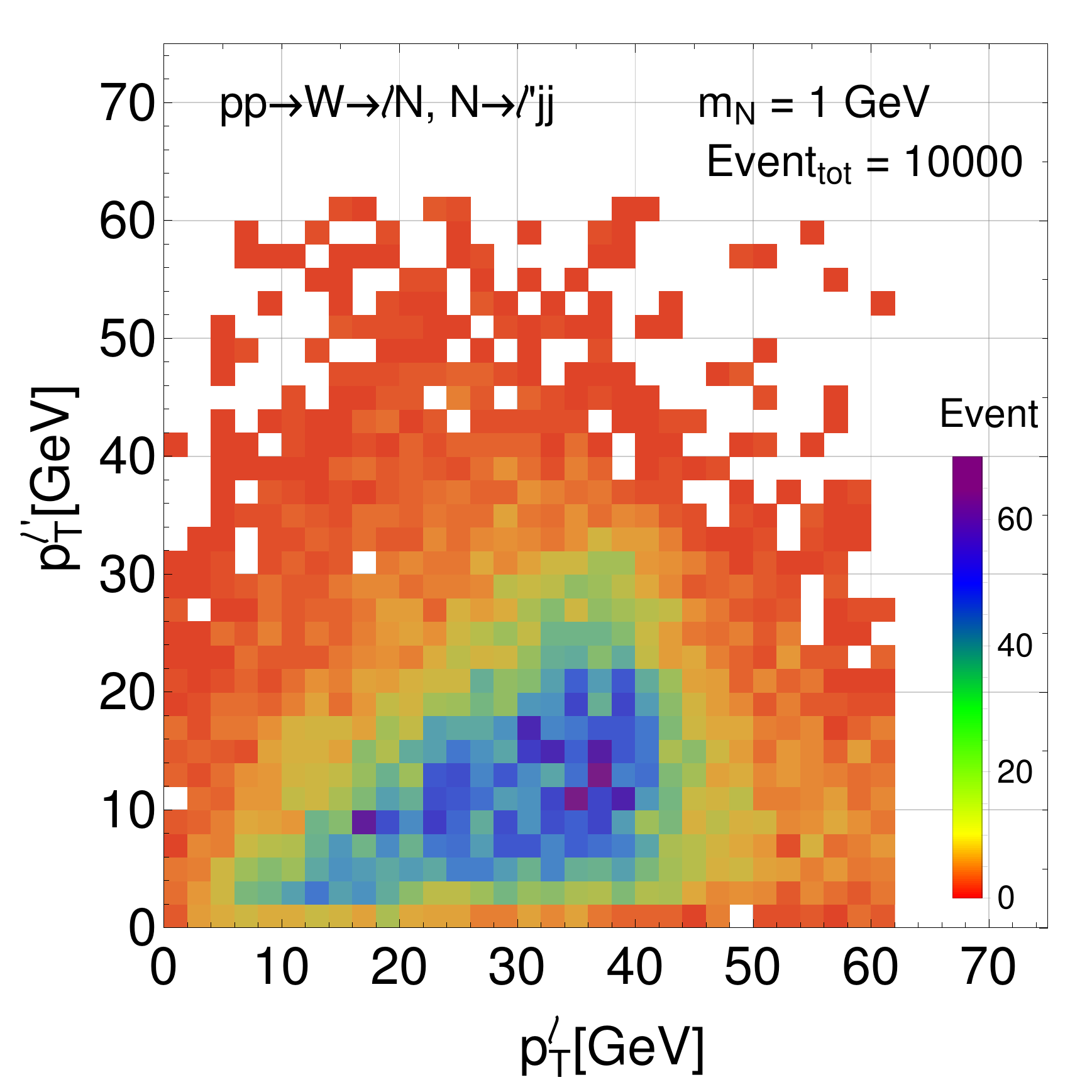}
    \includegraphics[width=0.45 \textwidth]{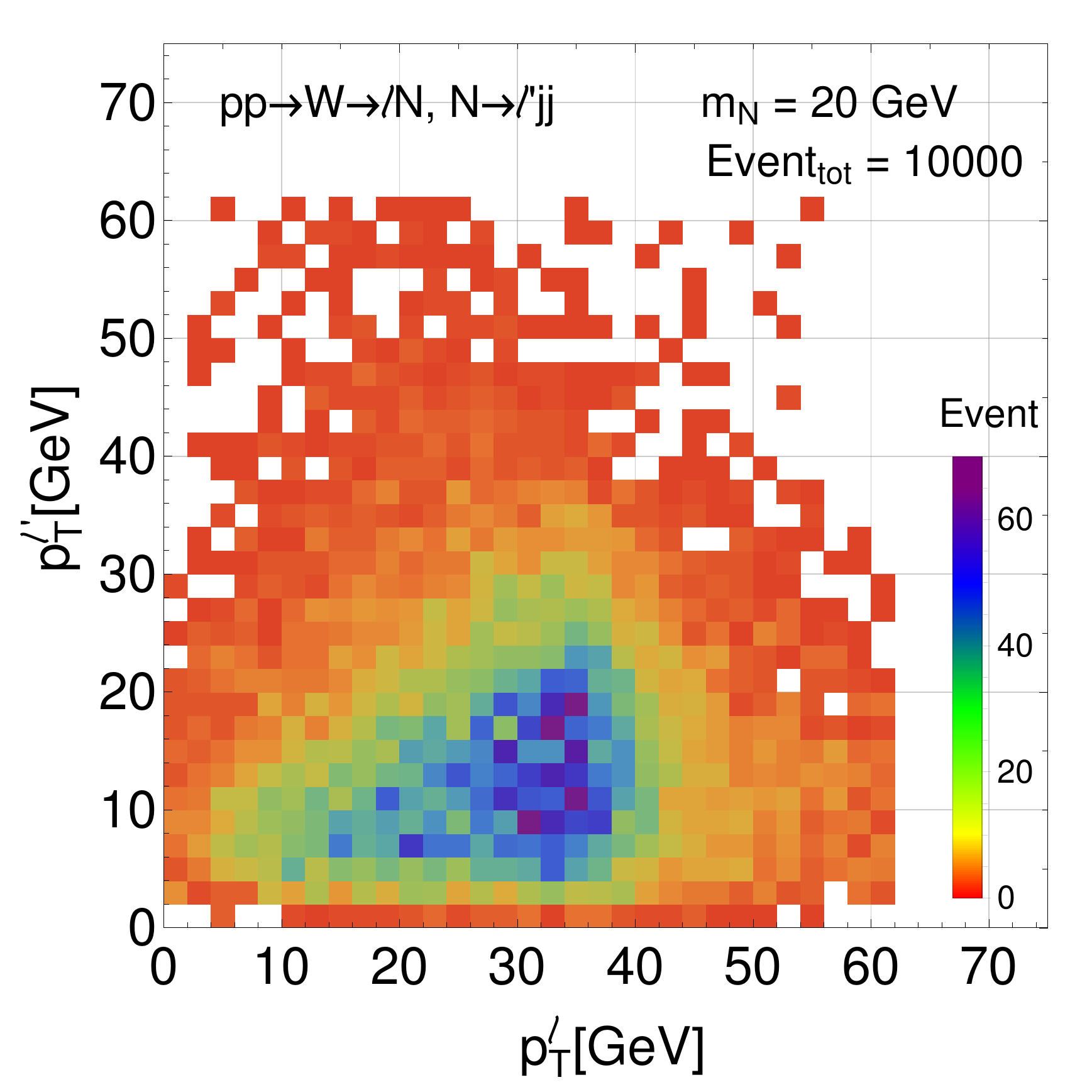}
    \caption{Kinematics of the signals from the process $p p \to W \to \ell N$ with subsequent decay $N \to \ell' q q'$.
        \textit{Top panel}: the transverse momentum $p_T^N$ versus the pseudo-rapidity $\eta_N$ of the sterile neutrino at the 13 TeV LHC. 
        \textit{Bottom panel}: the transverse momentum distribution of prompt lepton $\ell$ and the displaced lepton $\ell'$. 
        For the distributions, we have used mass $m_N = 1$~GeV and $20$~GeV shown in the left panels and right panels, respectively.
        The total number of events is 10000, which corresponds to $\sin^2\theta = 5.5\times 10^{-8}$ at integrated luminosity $\mathcal{L} = 3 ~{\rm ab}^{-1}$ for HL-LHC. 
    }
    \label{fig:wdecays}
\end{figure}

It is worth noting that the existence of two other SM venues for heavy sterile neutrino production, namely 
the $Z$ boson decay and the Higgs boson decay. These two channels will become competitive if displaced-track triggers becoming available \cite{Gershtein:2017tsv}, which can provide a similar amount of sterile neutrino in addition
to the $W$ boson decay. Furthermore, the Higgs channel has two exciting features, namely Higgs specific trigger options and large branching fractions. There are many sub-leading Higgs production channels which can be triggered on, especially the vector boson fusion channel and weak boson associated production channel. Although the Higgs boson production rate is three to four orders of magnitude smaller than the $W$ and $Z$ boson, the branching fraction of the Higgs boson to $\nu + N$ can be five orders of magnitude larger, given its small total width. 
In this study, we do not include these channels and save these interesting new production modes for future studies in association with displaced triggers.

On the top panel of Fig.~\ref{fig:wdecays}, we show the transverse momentum and pseudo-rapidity $p_T^{N}$--$\eta_N$ distribution for sterile neutrino $N$ from $W$ production 
$p p \to W \to \ell N$. The process is generated by {\tt MadGraph5\textunderscore aMC$@$NLO}~\cite{Alwall:2014hca}, and the parton shower is performed by {\tt Pythia8}~\cite{Sjostrand:2006za, Sjostrand:2007gs}. 
The $\eta$ distribution of $N$ is symmetric and dominantly within the range between $\left[-4, 4\right]$. Moreover, its transverse momentum $p_T^N$ peaks around 30$\sim$40~GeV, which is dictated by the maximum momentum it can obtain from $W$ decay in the center of mass frame, $\left( m_W^2 + m_N^2 \right)/\left(2 m_W \right)$. For the events with $p_T^N$ larger than this value, the initial state radiation provides additional transverse momentum of the $W$ boson system.
In the bottom panel of Fig.~\ref{fig:wdecays}, we show the $p_T$ distribution for both the prompt lepton $\ell$ from $W$ decay and displaced lepton $\ell'$ from $N$ decay.  
The distribution of prompt lepton $p_T^{\ell}$ is similar to that of $p_T^N$, with the difference
that the maximum momentum in center of mass frame becomes $\left( m_W^2 - m_N^2 \right)/\left(2 m_W \right)$. The displaced lepton transverse momentum, $p_T^{\ell'}$, is dominantly distributed
within $\left[5, 20 \right]$ GeV, because the other two particles in the $N$ decay take away approximately two thirds of the available energy. 

\section{New searches strategy for sterile neutrinos}
\label{sec:analysis}

There are many challenges to overcome to reach good sensitivities in the search for sterile neutrinos at the LHC. 
A major challenge is how to achieve a good signal selection efficiency with effective background suppression. Typical neutrino mass models point to tiny couplings between the sterile neutrino and the SM electroweak gauge bosons, leading to displaced decays of the sterile neutrinos.
Therefore, a very effective strategy is necessary to pick out these displaced events at the LHC. Another aspect is triggering. The sterile neutrino can be produced in electroweak processes, which typically give rise to soft objects. At the same time, due to the low signal rates, the triggering  needs to be as efficient as possible. 

We consider the process of $pp\to W \to N \ell  $, with a subsequent
displaced decay of the sterile neutrino $N \to \ell' j j$. Since the prompt lepton $\ell$ is hard, one can in principal trigger 
the event using single lepton trigger. According to CMS Phase-2 upgrade of the level-1 trigger  \cite{Collaboration:2283192}, the trigger thresholds (with track trigger) on a single lepton  is $p_T > 27~(31)$ GeV for isolated  (non-isolated) $e$,  and $18$ GeV for $\mu$.
As shown in the prompt lepton $p_T$ distribution in Fig.~\ref{fig:wdecays}, the signal events
will be reduced substantially by the single lepton trigger. Therefore, we consider double lepton trigger with the leading lepton 
$p_T> 19$ GeV and the sub-leading lepton with $p_T> 10.5$ GeV, also benefits from the track trigger \cite{Collaboration:2283192}. 
In reality, the combination of triggers will be used in the experiment. Therefore, it is expected that the trigger efficiency
will be better than using a single trigger category. Therefore, our approach on the triggering in this study is conservative.

\begin{figure}[htb]
    \centering
    \includegraphics[width=0.32 \textwidth]{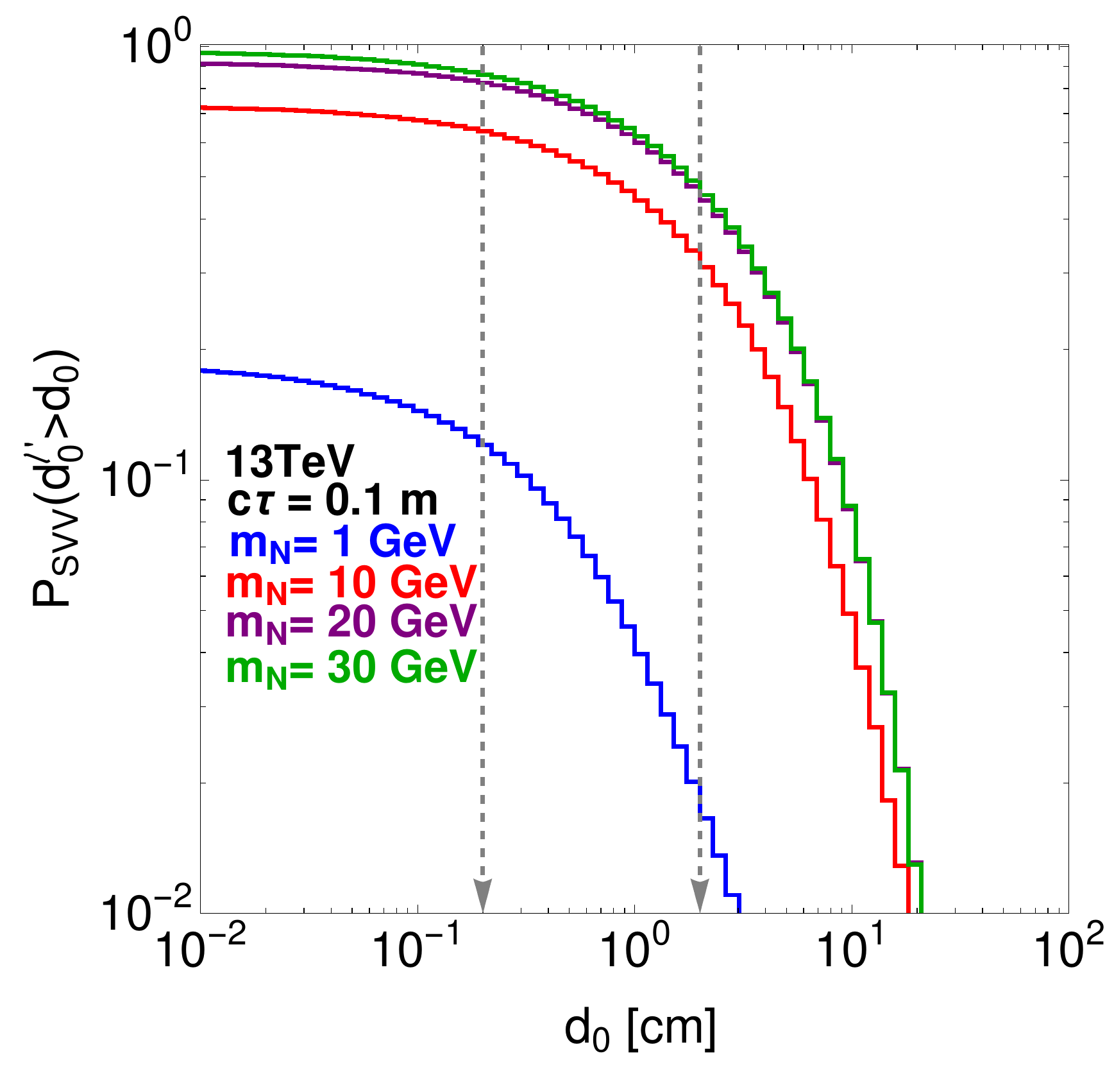} 
    \includegraphics[width=0.32 \textwidth]{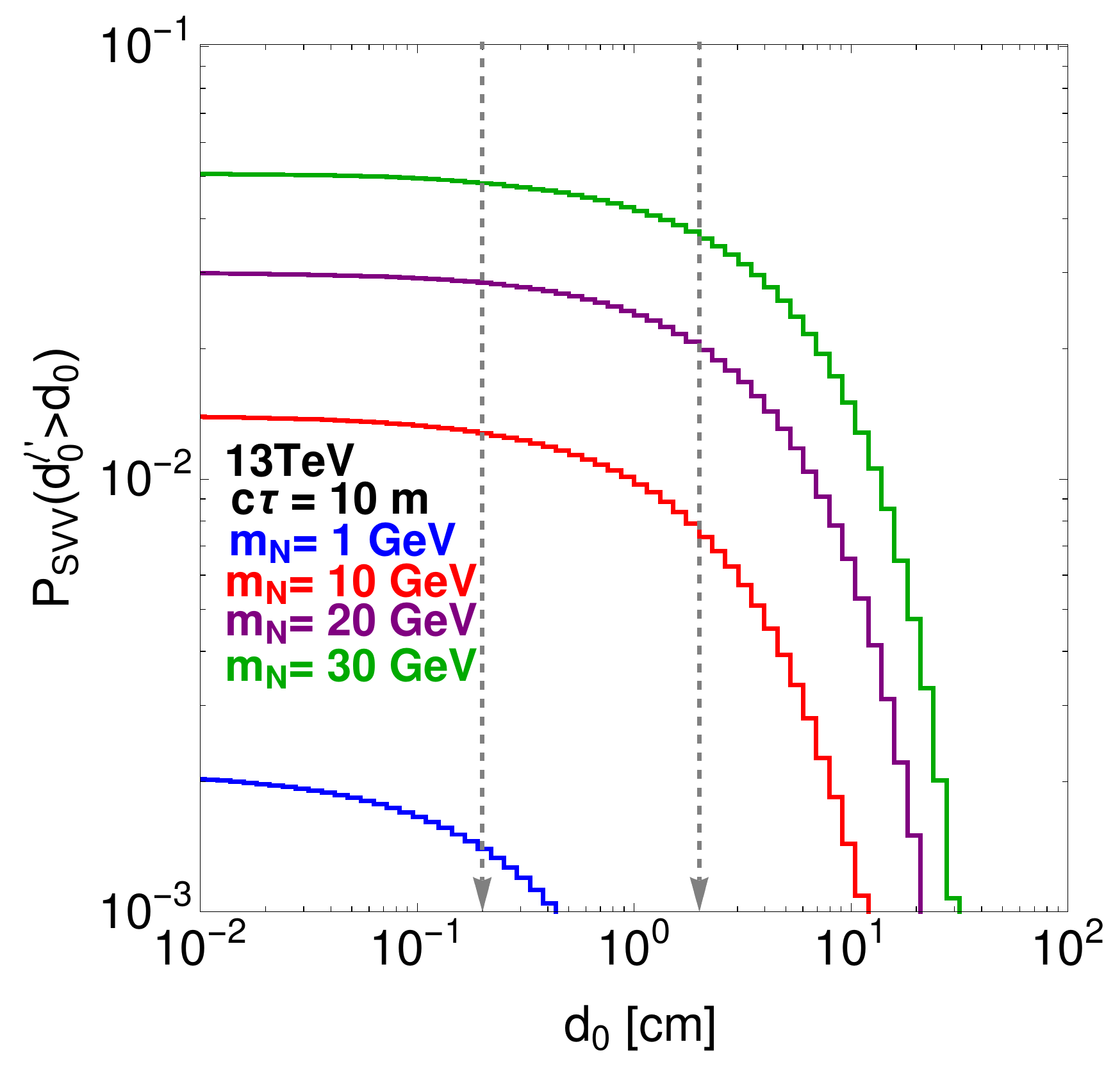}
    \includegraphics[width=0.32 \textwidth]{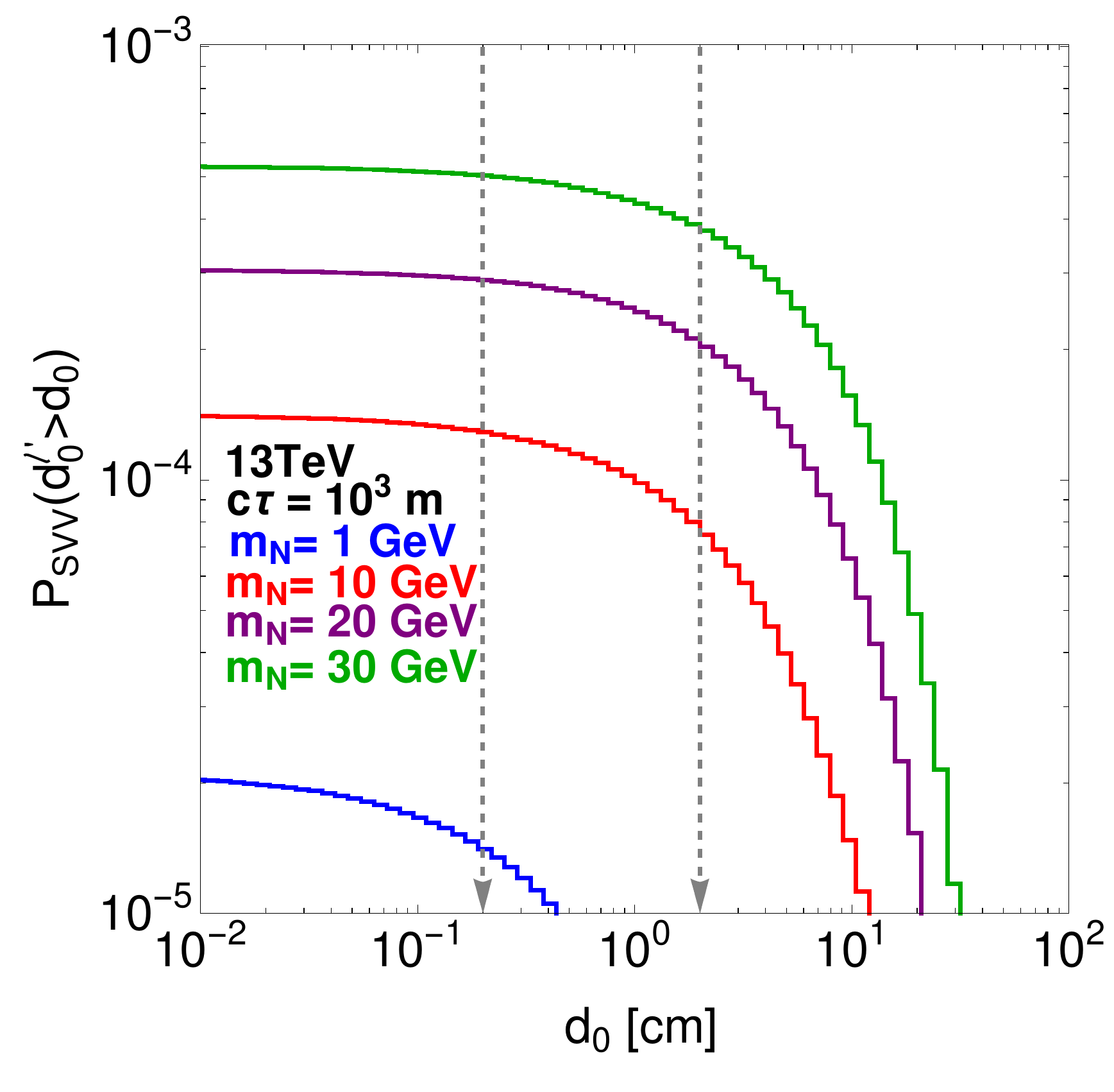}  
    \caption{The survival probability $P_{\rm SVV}(d_0^{\ell'} > d_0)$ for the lepton $\ell'$ from the sterile neutrino decay 
        $N \to \ell' j j$, as a function of the minimal $d_0$ cut.
    From left to right, the proper lifetime of sterile neutrino $c\tau$ are set as $0.1$, $10$ and $1000$ meters, respectively. The two gray dashed arrows show $d_0 = 0.2$~cm and $d_0 = 2$~cm, which correspond to our benchmark selection cut.
    }
    \label{fig:impactd0Distribution}
\end{figure}

For the off-line signal selection, the major advantage is the presence of a bundle of tracks with large (transverse) impact parameter, $d_0$. 
We denote the decay position of sterile neutrino $N$ as $(x ,y ,z)$ with the production point being 
the origin, and the momentum of a daughter particle from $N$ decay in x-y plane is denoted as $(p_x, p_y)$. We can define the transverse impact parameter $d_0$ as\footnote{For simplicity, we ignore the bending of tracks in this study as the displaced leptons $p_T$ is sufficiently large. Furthermore, the hard prompt lepton enable us to reconstruct the three-dimensional position of the primary vertex, additional requirement on the three-dimensional impact parameter can compensate this simplification and improve our results.}
\begin{align}
d_0=\sqrt{x^2+y^2-\frac{ \left( x  p_x + y p_y \right)^2 }{p_x^2+p_y^2}}.
\end{align}

We require the sterile neutrino $N$ to decay within the detector region  $r < 0.5$ m and 
$|z| < 1.2$ m, where $r$ is the radial distance to the beam line. These numbers are chosen to guarantee that the lepton from the decay of the sterile neutrino would pass at least four Outer Tracker layers in either central or forward region, based on the Phase-2 upgrade of the CMS tracker \cite{Collaboration:2272264}.
Moreover, each layer of Outer Tracker consists of two closely spaced silicon sensors, 
which are called $p_T$ modules to reject the low $p_T$ tracks. The threshold is about 2 GeV which is easy to satisfy for our signal, based on the $p_T$ distribution of displaced lepton in Fig.~\ref{fig:wdecays} and in particular for the events passing our trigger selection. The silicon sensors have a good granularity to provide sufficient spatial resolution
and the module has a $p_T$ resolution of $5\%$.
If the electromagnetism calorimetry also has an excellent pointing resolution, using one layer of the Outer Tracker could be good enough to identify the direction of the track and measure the $p_T$,  the detector region can be enlarged to $r \lesssim 1$ m and $|z| \lesssim 2.5$ m. In this case, the volume of the detector is increased by a factor of 8, leading to a significant improvement of the reach. We leave this possibility to future experimental studies and keep our conservative requirement on the number of track layers to hit.

The signal is selected first by imposing the  dilepton trigger with $|\eta_\ell| < 2.5$ and requiring the long-lived sterile neutrino $N$ decays in the detector region $r < 0.5$ m and 
$|z| < 1.2$ m. We further require the displaced lepton in the event  has a large transverse impact
parameter $d_0^{\ell' }  > d_0$.
The survival possibility is the number of such events divided by the total signal
events, which we denoted as $P_{\rm SVV}$. 
We plot its distributions with different sterile neutrino mass $m_N$ and proper lifetime $c\tau$ in Fig.~\ref{fig:impactd0Distribution}. For a given proper lifetime, a smaller sterile neutrino mass $m_N$ leads to a lower
$P_{\rm SVV}$. This is a result of the kinematics of the resonance decay.
Smaller sterile neutrino mass has larger boost factor, reducing the decay probability of sterile neutrino inside the detector region and reducing the displaced lepton opening angles to give a smaller $d_0$. 
For $c\tau = 0.1$ m, a significant portion of the sterile neutrino produced decay within the region of the detector we focused on. A notable exception is when the sterile neutrino is light. For example, the $P_{\rm SVV}$ is significantly smaller for $m_N=1$ GeV than other masses
 since it would often decay outside the region  due to the large boost. For 
$c\tau = 10$ m and $10^3$ m, the survival probabilities for different $m_N$s are roughly proportional to
$m_N^{-1}$ due to the boost factor of $N$. For the same $m_N$, the probabilities of
$c\tau = 10$ m and $10^3$ m  differ by  a factor of 100, exactly proportional to $(c\tau)^{-1}$ as expected in the long lifetime limit.
In summary, the survival probabilities in Fig.~\ref{fig:impactd0Distribution} are mostly determined by 
the probability for $N$ decay inside the required region, which is proportional to distance parameter $d_N^{-1}$. Here, $d_N = c\tau \gamma_N \beta_N$ is the expected decay distance in the laboratory 
frame, where $\gamma_N$ is the boost factor and $\beta_N$ is the velocity of $N$.

The signal contains one prompt lepton and another displaced lepton. The estimation of the Standard Model (SM) background 
of a newer search for a long-lived particle is always a challenge at the LHC.
Fortunately, many important features of the corresponding background have been effectively explored as the control regions in a search for displaced electron plus muon search at CMS~\cite{CMS-PAS-EXO-16-022}. This search was performed with $2.6~\ifb$ of 13 TeV LHC data, and it looks for a pair of displaced isolated leptons with different flavors with minimum $p_T$ of 38 GeV each, targeting signals from pair produced top squarks decaying leptonically. The control regions (CR) of the background for this search helps to identify the dominant background of our new signals for long-lived sterile neutrinos. Specifically, the CR-III and CR-IV of this analysis requires one prompt lepton with transverse impact parameter {\it smaller} than 200~$\mu$m and a displaced lepton with transverse impact parameter {\it greater} than 100~$\mu$m. The CR-III looks for a heavy flavor jet plus a displaced electron and CR-IV looks for a heavy flavor jet plus a displaced muon. CR-III is triggered by a singlet electron trigger (with electron $p_T>20$~GeV), while CR-IV is triggered by a muon plus jet trigger (with a muon $p_T>10$~GeV and jet $p_T>12$ GeV), both with low $p_T$ threshold for the leptons~\cite{Khachatryan:2016bia}. 

The CMS study~\cite{CMS-PAS-EXO-16-022} shows that the displaced leptons dominantly come from the heavy flavor QCD events (HF), because B and D mesons have sizable lifetimes.
They simulate the HF+$\ell$ data, requiring one tagged b-jet and one displaced lepton from the other 
heavy flavor quark. 
They further use a data driven method with the $e+\mu$ data and obtain the
$d_0$ spectrum for one lepton while requiring the other lepton being prompt ($d_0 < 200 $ $\mu$m). 
In figure 3, they show the agreement in the $d_0$ distribution between HF+$\ell$ and $e+\mu$ data
\footnote{
However, there are some differences between the CR-III (IV) data and 
the HF+$\ell$ data. First, the CR-III and CR-IV do require one prompt and one displaced
leptons, and the leptons satisfy the preselection for leptons. Moreover, the heavy flavor jet is not required for CR-III
and IV, which is required for HF+$\ell$ control data only. Second, HF+$\ell$ control data only requires one displaced lepton from the heavy flavor quark. 
Thus, HF+$\ell$ control data is not the same as CR-III and IV data, and is used to provide the $d_0$ shape information only.
}.
As a result, the detailed studies of these control regions in Ref.~\cite{CMS-PAS-EXO-16-022} confirm the following crucial fact about the displaced background distribution. First of all, the displaced leptons are dominantly from heavy flavor jets (b-jets) by using the ``tag and probe'' method where the jet recoil against the displaced lepton is tagged as a b-jet. The subleading background is from $t\bar t$ which is smaller by more than one order of magnitude, agreeing well with our simulation. Furthermore, for the displaced lepton, the (normalized) differential distribution as a function of the transverse impact parameter is shown to be the same for {\it isolated} and {\it non-isolated} leptons.
%\footnote{CR-III and CR-IV are dominated by heavy flavor plus lepton background. When describing the properties, we use them interchangeably.}.

In Fig.~2 of Ref.~\cite{CMS-PAS-EXO-16-022}, the background events are dominated by heavy flavor. 
Summing up all the $d_0$ bins,
the corresponding cross-sections are 16.6 pb for HF$+e$ and 259 pb for HF$+\mu$ respectively. We denote them as $\sigma_{{\rm HF}+e}^{\rm CMS}$ and $\sigma_{{\rm HF}+\mu}^{\rm CMS}$ respectively. 
The HF$+\mu$ background is much larger than HF$+e$ background because the muon plus jet trigger requires a softer lepton than the single electron trigger.
We used these results to validate our own simulation. 
In particular, we generated $b\bar{b}$ events and require $e$ or $\mu$ to show up in the event after 
hadron fragmentation using {\tt Pythia8}. 
We apply the CR-III and CR-IV triggers separately and further require at least one b-tagged jet.
For HF$+e$, the isolation requirement for the electron is $\Delta R_e < 0.3$ and the additional $p_T$ sum within the isolation cone should be less than $3.5\%$ ($6.5\%$) of the electron's $p_T$ in the barrel (endcap) region. For HF$+\mu$, the isolation requirement for the muon is $\Delta R_\mu < 0.4$ and the additional $p_T$ sum within the isolation cone should be less than $15\%$ of the muon $p_T$. We implement these requirements by modifying the {\tt Delphes3} \cite{deFavereau:2013fsa}. For b-tagging efficiency, we have used the working point in \cite{Chatrchyan:2012jua}, with $55\%$ tagging efficiency for a b-jet with $p_T > 30$ GeV and 
$|\eta|< 2.4$. We found the corresponding cross-sections to be 21.9 pb and 241.2 pb for HF$+e$ and HF$+\mu$ backgrounds, which are consistent with our extraction from Ref.~\cite{CMS-PAS-EXO-16-022}. We denote them as 
$\sigma_{b\bar{b}(e)}^{\rm icut}$ and $\sigma_{b\bar{b}(\mu)}^{\rm icut}$ which are the cross-sections after cuts with isolation requirement on leptons, denoted as ``icut'' in the superscript. 
Such consistency provides confidence when we calculate the background to our new search\footnote{We thank Bingxuan Liu on the analysis of Ref.~\cite{CMS-PAS-EXO-16-022} for clarifying the details of the control region in the analysis and confirming our scaling method.}.

\begin{figure}[tb]
	\centering
	\begin{tabular}{cc}
		\includegraphics[width=0.35 \textwidth]{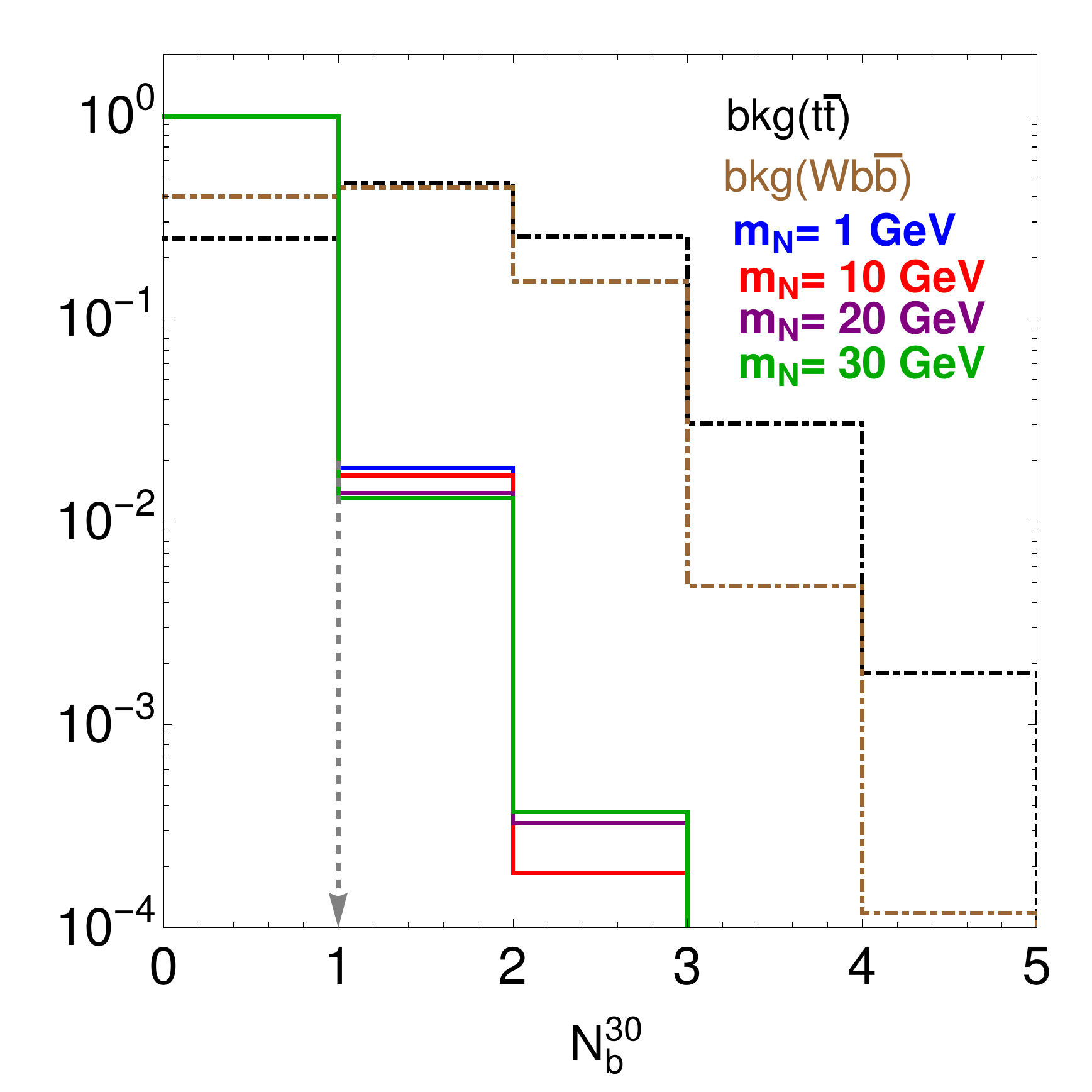} &
		\includegraphics[width=0.35 \textwidth]{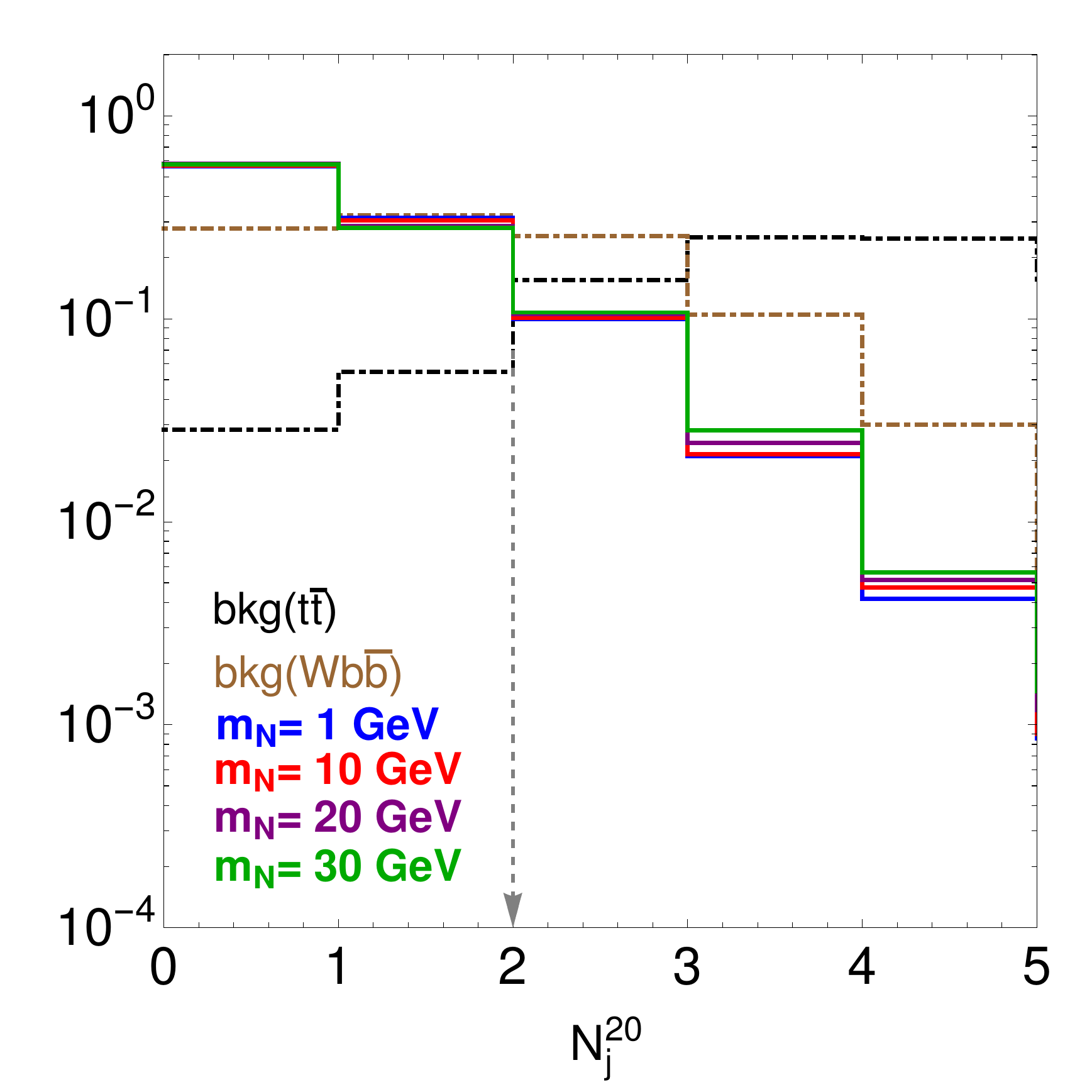} \\
		(a) & (b) \\
		\includegraphics[width=0.35 \textwidth]{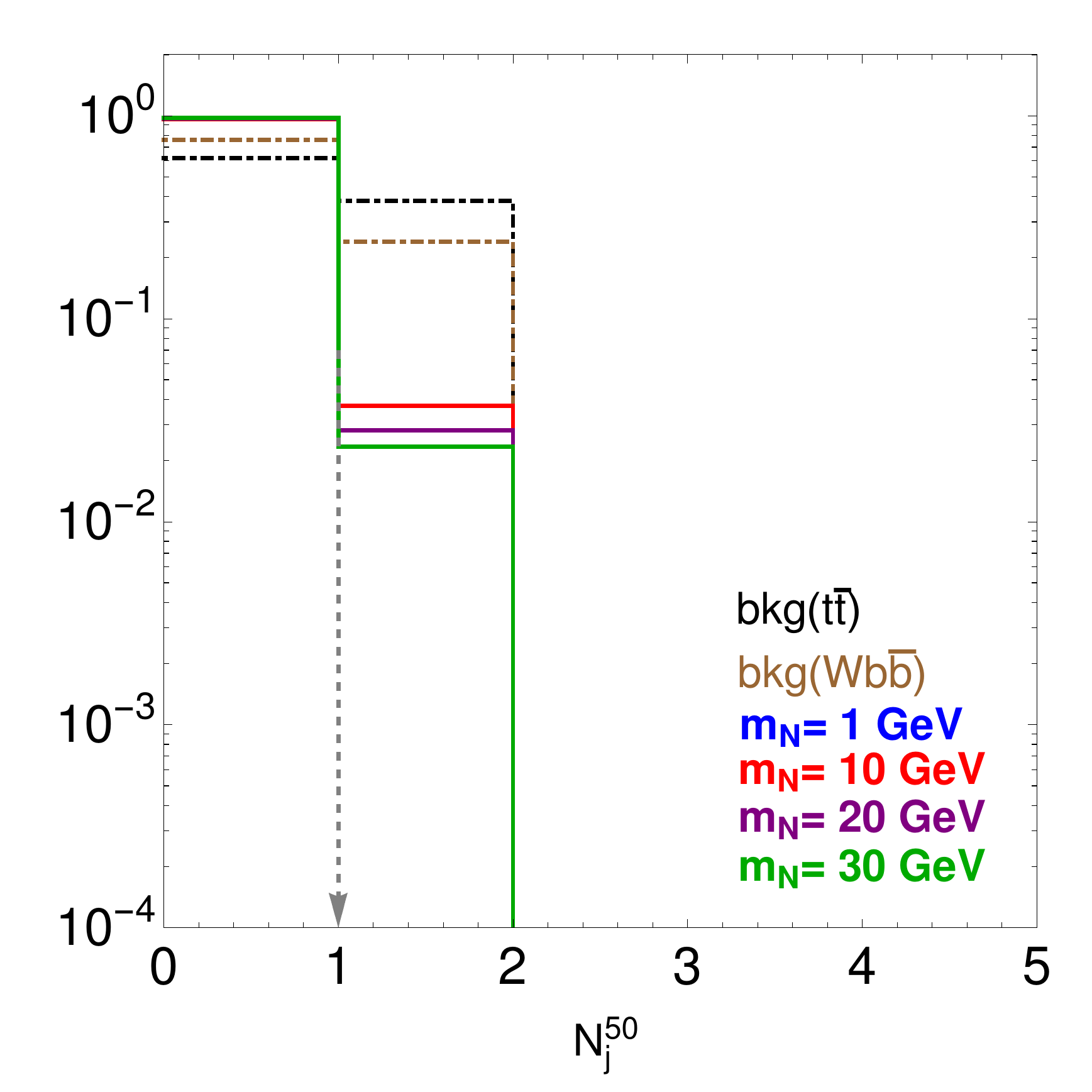} &
		\includegraphics[width=0.35 \textwidth]{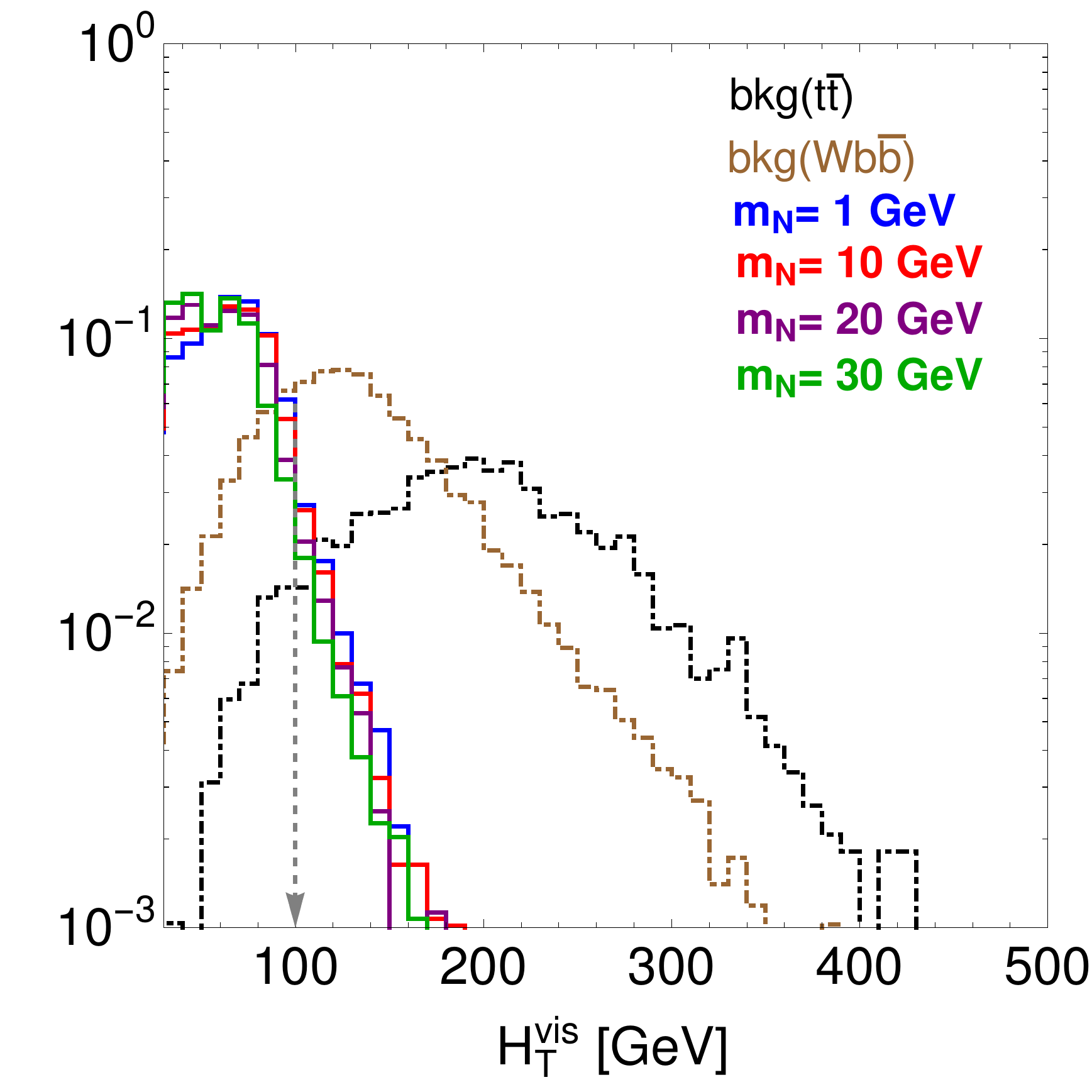} \\
		(c) &(d)
	\end{tabular}
	\caption{The normalized distribution for the number of b-jets with $p_T^b > 30$ GeV denoted as $N_b^{30}$, the number of jets with $p_T^j > 20$ GeV denoted as $N_j^{20}$, the number of jets with $p_T^j > 50$ GeV denoted as $N_j^{50}$  and $H_T^{\rm vis}$, where $H_T^{\rm vis}$ is the scalar sum of $p_T$ for all visible objects, including hadronic jets and leptons. 
		The two SM backgrounds 
		$t\bar t\to b\bar b+\ell+X$ and $W+b\bar{b}, ~W\to \ell \nu$ are dot-dashed lines with black and brown color respectively. The signals are solid lines with blue, red, purple and green colors for
		increasing $m_N$. The dashed gray lines with arrow indicates the optimization cuts that the
		events to its left are retained, namely $N_b^{30} = 0$, $N_j^{20} < 2$, $N_j^{50} =0$ and $H_T^{\rm vis} < 100$ GeV. The above cuts have been applied after its distribution have been shown in figure (a), (b), (c), and (d).
	}
	\label{fig:vardistribution}
\end{figure}

The  displaced lepton in the signal is often {\it non-isolated}. Therefore, a major background comes from  events with a displaced {\it non-isolated} lepton from heavy flavor jets and a prompt lepton. Consequently, the leading backgrounds for our signal could be $W+b\bar b$ with $W$ decaying leptonically, and 
$t\bar t$ with one of the top quarks decay leptonically. 
The number of the background can be calculated as
\begin{align}
N_{\rm{bkg}}= \frac{\sigma_{{\rm HF}+e}^{\rm CMS} + \sigma_{{\rm HF}+\mu}^{\rm CMS} }{\sigma_{b\bar{b}(e)}^{\rm icut} + \sigma_{b\bar{b}(\mu)}^{\rm icut} } \left( 
\sigma_{W+b\bar b, W\to \ell\nu}^{\rm ncut} \times \epsilon_{\rm opt}^{W+b\bar{b}}+\sigma_{t\bar t\to b\bar b+\ell+X}^{\rm ncut} \times \epsilon_{\rm opt}^{t\bar{t}} \right) \times  \mathcal{L}_{\rm HL-LHC},
\label{eq:Nbkg}
\end{align}
where the ``ncut'' in the upper script means requiring jets with $p_T^j > 20$ GeV while having one \textit{non-isolated} lepton in the final states. In the event generation, we require b quark $p_T^b > 30$ GeV at parton level to ensure an energetic \textit{non-isolated} lepton from its hadronic fragmentation. Otherwise,
it is difficult to pass the lepton $p_T$ cut.
The lepton $\ell$ represent both $e$ and $\mu$. Further optimization selections are applied to both the signal and background, whose selection efficiencies are denoted as $\epsilon_{\rm opt}$ in Eq.~\ref{eq:Nbkg}.
The cross-sections $\sigma_{t\bar t\to b\bar b+\ell+X}^{\rm ncut}$ and $\sigma_{W+b\bar b, W\to \ell\nu}^{\rm ncut} $ are found to be 136 pb and 3.8 pb respectively, after applying  the ``ncut''.

To further reduce the SM background, especially those from the $t\bar t$ process, additional cuts on the hadronic activities can help. In Fig.~\ref{fig:vardistribution}, we show the normalized distribution for the number of b-jets with $p_T^b > 30$ GeV denoted as $N_b^{30}$, the number of jets with $p_T^j > 20$ GeV denoted as $N_j^{20}$, the number of jets with $p_T^j > 50$ GeV denoted as $N_j^{50}$  and $H_T^{\rm vis}$, where $H_T^{\rm vis}$ is the scalar sum of $p_T$ for all visible objects, including hadronic jets and leptons. 
We choose the optimization condition $N_b^{30} = 0$,
$N_j^{20} < 2$, $N_j^{50} = 0$ and $H_T^{\rm vis} < 100$ GeV to suppress the SM background without a significant reduction of the signal events. In Fig.~\ref{fig:vardistribution}, the above cuts have been applied after its distribution have been shown in sub-figures (a), (b), (c), and (d).
We have included the jet matching in the background simulation. For $N_{j, b}$ and $H_T$ observables,
the $t \bar{t}$ background has slight differences between the results with and without jet matching.
In addition, the $W+b\bar{b}$ backgrounds have even smaller difference than $t\bar{t}$ background.
We also include the efficiency information for the dilepton trigger requirement $p_T^{\ell_1} > 19$ GeV and $p_T^{\ell_2} > 10.5$ GeV. 
The effects of various selection cuts and the total efficiencies are given in Table.~\ref{table:cuts}.

\begin{table}[!h]
    \begin{center}
        \begin{adjustbox}{max width=\textwidth}
            \begin{tabular}{|c|c|c|c|c|c|c|c|c|}
                \hline
                Efficiency & $\sigma^{\rm ncut}$ (pb)& $N_{ b}^{30} =0$ & $N_{j}^{20} <2$& $N_{j}^{50} =0$ & $H_T^{\rm vis} < 100 $ GeV  & $p_T^{\ell_1} > 19 \rm GeV$ & $p_T^{\ell_2} > 10.5 \rm GeV$&$\epsilon_{\rm opt}$  \\
                \hline
                $t\bar t\to b\bar b+\ell+X$ & 136 &  0.25 & 0.08& 0.62 &0.43 & 0.055 & 0.42 &  $1.2 \times 10^{-4}$\\
                \hline
                $W+b\bar b, W\to \ell\nu$ & 3.8 &   0.40 & 0.60 & 0.76& 0.40 & 0.27 & 0.29 & $5.7 \times 10^{-3}$ \\
                \hline
            \end{tabular}
        \end{adjustbox}
    \end{center}
    \caption{The selection efficiency for the two dominant SM backgrounds $t\bar{t}$ with only one top quark decaying leptonically and $W+ b\bar{b}$ with $W$ decaying leptonically. The efficiencies reported in the table are the additional suppression with respect to all the previous cuts, and $\epsilon_{\rm opt}$ is the total efficiency given by  the product of all the cut efficiencies. 
    }
    \label{table:cuts}
\end{table}

After combining all the cuts and optimizations, the backgrounds from $t\bar{t}$ and $W+b\bar{b}$  are 0.017 pb and 0.022 pb respectively, which are comparable in size. Multiplying by the
integrated luminosity $3000 ~{\rm fb}^{-1}$ from HL-LHC, the corresponding number of SM backgrounds are about 51000 and 65100 respectively. 
We assume for the SM background, the $d_0$ distribution for the lepton does not depend strongly on the $p_T$ of the lepton and other optimization cuts on jets. 
Together with the fact that \textit{isolated} and \textit{non-isolated} lepton has the same distribution, we can use the normalized differential $d_0$ distribution from Ref.~\cite{CMS-PAS-EXO-16-022} to determine the background. 

\begin{figure}[tb]
    \centering
    \includegraphics[width=0.45 \textwidth]{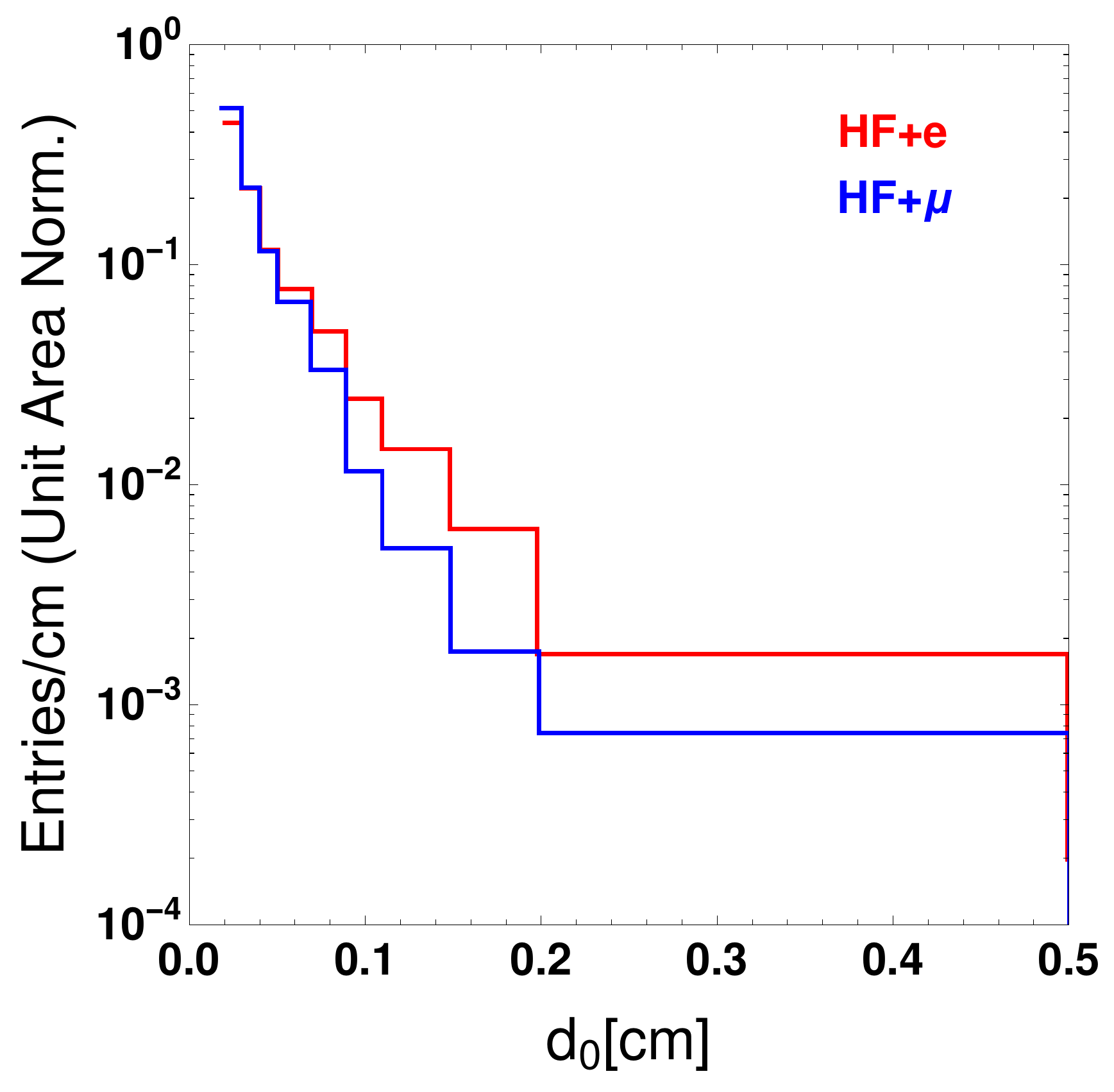} 
    \includegraphics[width=0.45 \textwidth]{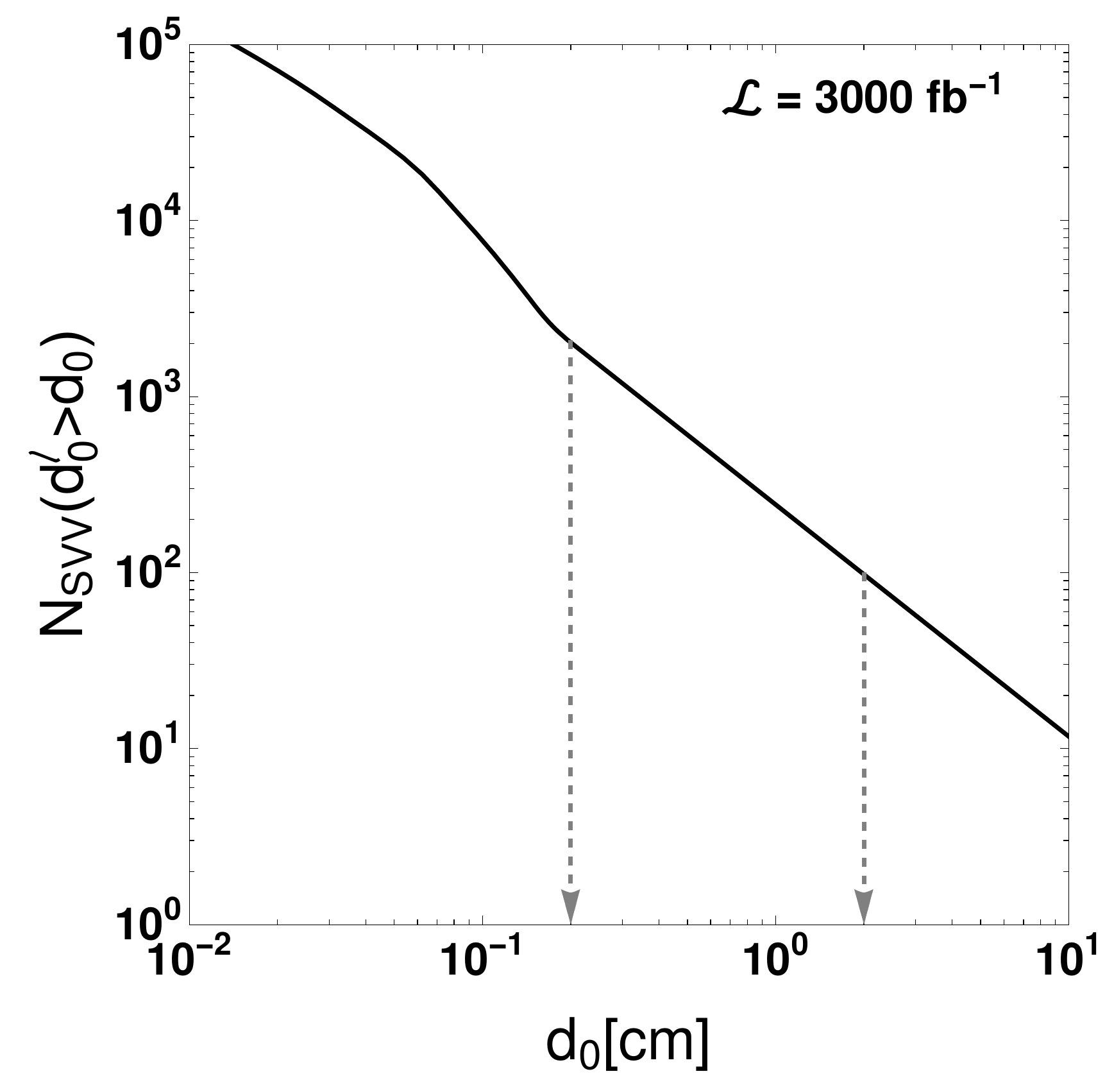} 
    \caption{\textit{Left panel}: The normalized $d_0$ distribution for HF$+e$ and HF$+\mu$ control regions from Ref.~\cite{CMS-PAS-EXO-16-022}.
    \textit{Right panel}: The total number of SM background after applying minimal $d_0$ cut with all the trigger and selection cuts applied, denoted as $N_{\rm SVV}\left(d_0^\ell > d_0 \right)$.
    The two dashed gray lines with arrow indicates two of our benchmark minimal $d_0$ cuts of 0.2~cm and 2~cm selection. 
    }
    \label{fig:d0bkgafterOPT}
\end{figure}

In Fig.~\ref{fig:d0bkgafterOPT}, we plot the total number of SM background after applying $d_0^{\ell} > d_0$ with all the cuts and optimizations applied. Since the differential distribution of $d_0$ from
CMS study \cite{CMS-PAS-EXO-16-022} stops at 0.5 cm, we linearly extrapolate their data beyond that for the case of a larger $d_0$ cut of 2~cm. Moreover, their last bin, ([0.2, 0.5] cm), contains the overflow entries, thus our extrapolation is reasonably conservative. 
The SM background decreases linearly with $d_0$ cut in Fig.~\ref{fig:d0bkgafterOPT}. 
After comparing with the $d_0$
accumulative distribution for signal in Fig.~\ref{fig:impactd0Distribution}, we adopt two benchmarks
for the displacement, requiring  $d_0 > $  0.2 cm and 2 cm, respectively.
From Fig.~\ref{fig:d0bkgafterOPT}, the total number of background event is around 2000 for $d_0> 0.2$ cm and 100 for $d_0 > 2$ cm. 
Despite the analysis of SM background above, it can also be estimated in a data-driven method similar to the CMS search. For instance, one can study the $d_0$ distribution of the non-isolated leptons using the ``tag and probe'' method, same as what have done in the heavy flavor plus lepton control region. Furthermore, one can also study the invariant mass distribution of the non-isolated lepton (plus hadrons) system. Future studies could also include more exclusive decays of the sterile neutrino to reconstruct the mass and reject the background more efficiently. Moreover, heavy long-lived sterile neutrinos will also be time-delayed. These additional features can help define the control region in a more sophisticated manner and will certainly improve the sensitivity.

With the above background estimation, we derive the $95\%$ C.L. sensitivity for sterile neutrino in red shaded region in Fig.~\ref{fig:NfromW}, where the left and right panel are for $d_0 > 0.2$~cm and $d_0 > 2$~cm, respectively. In Ref. \cite{CMS-PAS-EXO-16-022}, the systematic uncertainties for different SM backgrounds
vary from $5\%$ to $10\%$. Therefore, we conservatively assume systematic uncertainty for SM backgrounds to be $10\%$. The corresponding sensitivity curves with the systematic uncertainty are plotted in dashed red lines in Fig.~\ref{fig:NfromW}.  
In general, the upper edge of the sensitivity region corresponds to the shorter lifetime. It is driven by the lower cut on the transverse impact parameter $d_0$ and we see clear advantage for smaller cut on $d_0$, even though the number of SM background is larger. On the other hand, the lower edge of the red shaded region corresponds to the longer lifetime, which is insensitive to the $d_0$ cut. 

In Fig.~\ref{fig:NfromW}, we plot the leading constraints from CMS \cite{Sirunyan:2018mtv} and DELPHI \cite{Abreu:1996pa}. The CMS collaboration \cite{Sirunyan:2018mtv} looks for heavy sterile neutrinos in events with three prompt charged leptons, while the DELPHI collaboration \cite{Abreu:1996pa} looks for both short-lived and long-lived sterile neutrinos in hadronic Z decay events. We have combined these two constraints together and shown the exclusions in shaded gray region. For sterile neutrino mass smaller than 2 GeV, the more
important constraints are from beam dump experiments, like NuTeV \cite{Vaitaitis:1999wq}, CHARM \cite{Bergsma:1985is, Vilain:1994vg}, BEBC \cite{CooperSarkar:1985nh}, and FMMF \cite{Gallas:1994xp}. Their constraints are also shaded in gray color. 
In Fig.~\ref{fig:NfromW}, we also show the projected sensitivity from a proposed displaced vertex search at the LHC in dot-dashed curves~\cite{Cottin:2018nms} and various proposals for fix target or satellite detector experiments~\cite{Adams:2013qkq, Alekhin:2015byh, Anelli:2015pba, Curtin:2018mvb, Kling:2018wct} in dashed curves. The LBNE collaboration \cite{Adams:2013qkq} is a long-baseline neutrino experiment also named as DUNE, which measures the sterile neutrino and active neutrino mixing by comparing the neutral current events between the near and far detectors. The SHiP collaboration \cite{Alekhin:2015byh, Anelli:2015pba} is a fixed target
facility (proton beam dump) probing the mixing by the signature of displaced secondary vertex from sterile neutrino
decay. The MATHUSLA collaboration \cite{Curtin:2018mvb} is a far detector proposal for the LHC and also measures the mixing with displaced sterile neutrino decay. The FASER \cite{Kling:2018wct} is another far detector proposal, but is placed in the forward region. Those future projections are plotted in dashed lines in the figure. It is clear that our more inclusive search allows us to explore parameter spaces of the sterile neutrino with lighter masses. Furthermore, when compared with other fixed target experiment and satellite detector experiments, our proposal covers new regions of sterile neutrinos whose lifetime is too short to have sufficient flux for these experiments. In the long run, if any discoveries were made, the regions overlapping with different searches and experiments will provide valuable information for the underlying new physics model.\footnote{A recent study \cite{Cottin:2018nms} projected that with a prompt lepton trigger with associated hadronic displaced vertex  at HL-LHC,  and assuming zero background, could cover the regime of sterile neutrinos with mass between 10-20 GeV, complementing our search.}

\begin{figure}[tb]
    \centering
    \includegraphics[width=0.46 \textwidth]{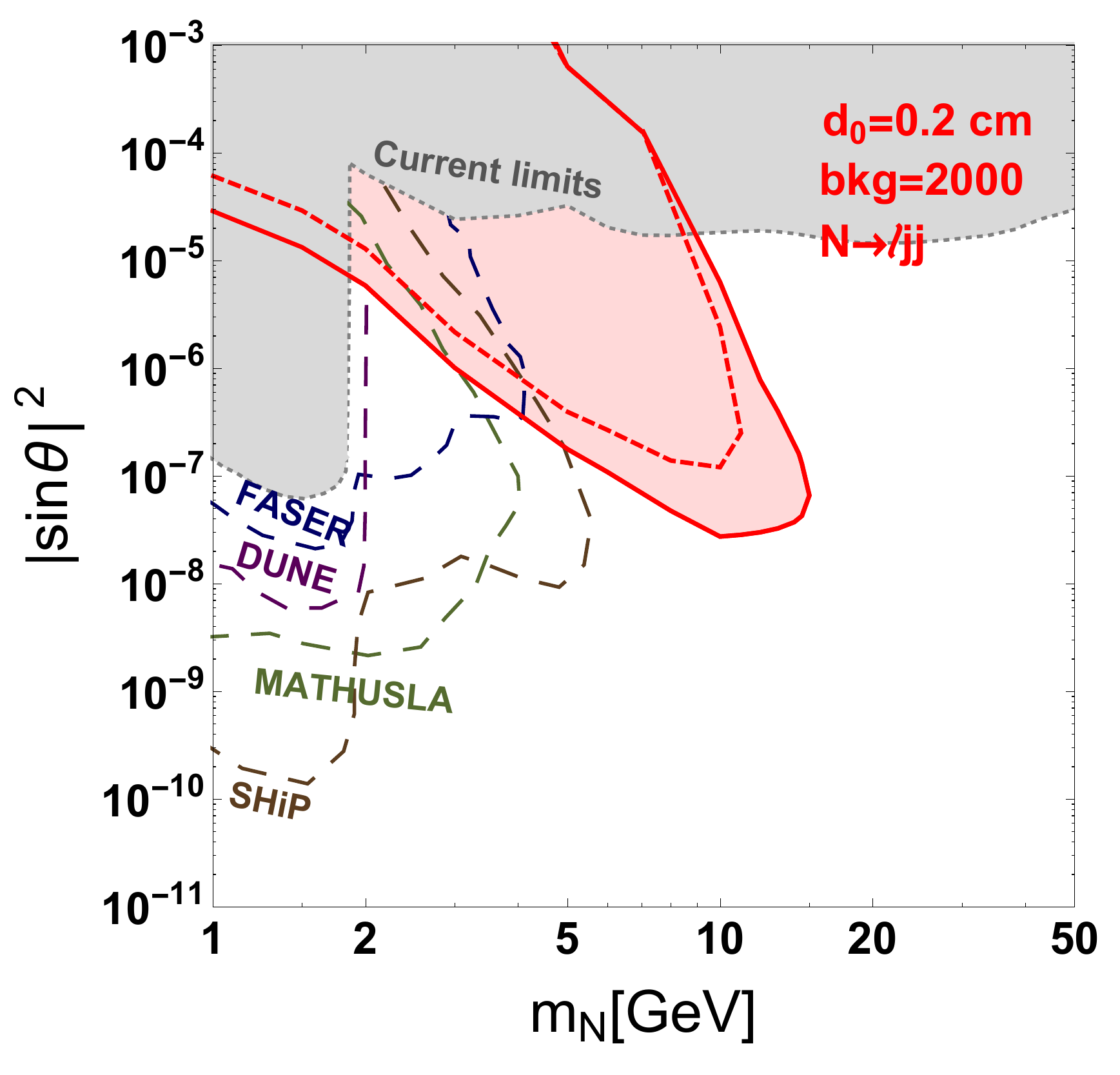} 
    \includegraphics[width=0.46 \textwidth]{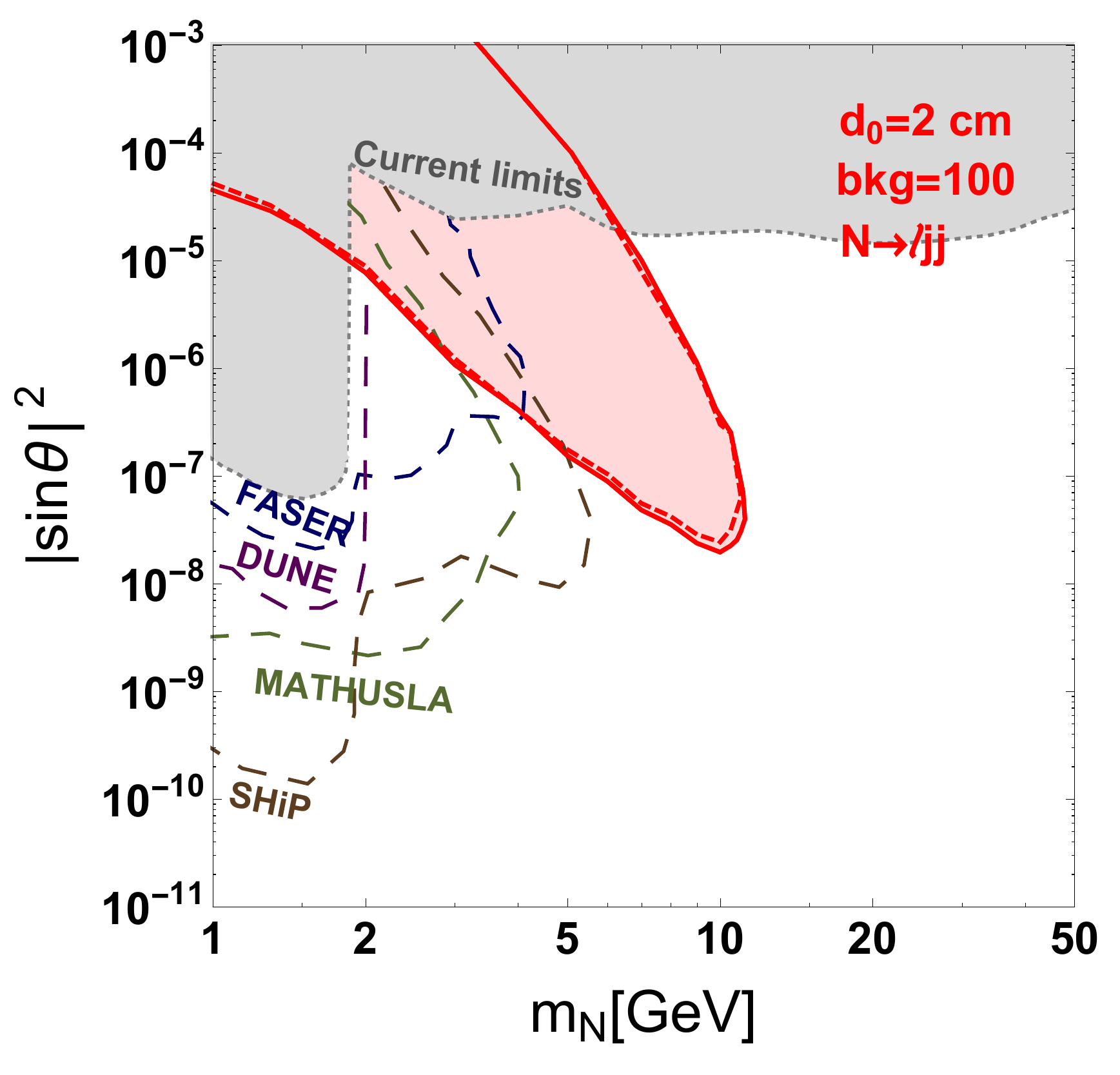}
    \caption{The $95\%$ C.L. reach for sterile neutrino from W gauge boson decay is plotted in the $m_N$-$\sin^2\theta$ plane with solid red lines. The dashed red lines have included $10\%$ systematic uncertainty effect. 
The existing constraints are from CMS \cite{Sirunyan:2018mtv} and DELPHI \cite{Abreu:1996pa} for mass larger than 1 GeV. While for mass smaller than $2$ GeV, the stronger constraints are from beam dump experiments like NuTeV \cite{Vaitaitis:1999wq}, CHARM \cite{Bergsma:1985is, Vilain:1994vg}, BEBC \cite{CooperSarkar:1985nh}, and FMMF \cite{Gallas:1994xp}. 
The existing current limits are shaded in gray color and labeled as ``Current limits".
The proposed sensitivity reaches for MATHUSLA \cite{Curtin:2018mvb}, FASER \cite{Kling:2018wct}, DUNE \cite{Adams:2013qkq} and SHiP \cite{Alekhin:2015byh, Anelli:2015pba} are shown in dashed curves. 
  }
    \label{fig:NfromW}
\end{figure}

\section{Conclusion and outlook}
\label{sec:conclusion}

In this paper, we study the signal of long-lived sterile neutrino $N$ from $W$-boson decay, $W \to \ell N$, with the subsequent decay of $N \to \ell' j j$. The
characteristic feature is a hard prompt lepton with a displaced lepton with large transverse impact parameter $d_0$. We neither reconstruct the displaced vertex nor cut on its invariant mass, therefore it can be sensitive for very low sterile neutrino mass. However, there is a crucial subtlety that with such small masses, the displaced lepton is usually \textit{non-isolated} from the other two jets in the same $N$ decay. 
To estimate the background, we have used the information from a search for displaced electron plus muon search at CMS~\cite{CMS-PAS-EXO-16-022} which studied relevant background in its control regions. It shows that for \textit{non-isolated} lepton, those from heavy flavor quarks are the dominant SM background.
Moreover, it demonstrates the important fact that the normalized $d_0$ differential distribution has
the same shape for \textit{isolated} and \textit{non-isolated} leptons. Therefore, we can use their
$d_0$ distribution for the \textit{non-isolated} lepton from heavy flavor quark background. We recast their control region selection and found a good agreement with their observations. This ensures our background estimations are  robust.
After proposing the optimization cuts for the signal,
we obtain the result for the sensitivity on parameters for the sterile neutrino. Our results can explore
the region with light $m_N$ and relative large mixing angle, which is not covered by the displaced vertex searches at LHC, beam dump and far detectors experiments.

There are many new exciting opportunities at the LHC, such as displaced tracker triggers proposed to be implemented at low-level at CMS by Ref.~\cite{CMS-PAS-FTR-18-018}. This trigger can be handy as a general trigger for long-lived particles in a broad class of theories~\cite{Alimena:2019zri}, especially for physics signals hard to be triggered on using traditional trigger.  
Although we have applied two lepton trigger in this study, adding the displaced tracker trigger into
the trigger menu can effectively lower the $p_T$ requirement on the leptons. Moreover, the Phase-2
upgrade of the tracker system of CMS contains the Outer Tracker layer which is consisted of two
closely spaced silicon sensors. It effectively doubles the hits of the track; therefore one can lower
the requirement on the number of penetrating layers,  which can increase the volume of the long-lived particle decay.
It has the potential to significantly increase the sensitivity but the detailed realization needs further studies. Moreover, for heavier sterile neutrinos above 10~GeV, the long-lived sterile neutrino are time-delayed with respect to the SM backgrounds. This feature can be further utilized for trigger and background suppression considerations~\cite{Liu:2018wte}.

\textbf{\textit{Note added.}} While we were finalizing this paper, Ref.~\cite{Dib:2019ztn}, \cite{Drewes:2019fou} and \cite{Bondarenko:2019tss} appeared on the arXiv studying similar topics. The main focus of Ref.~\cite{Dib:2019ztn} is looking for sterile neutrino decay to lepton plus pions, using displaced vertex associated with low $p_T$ muon. And their mass coverage for sterile neutrino is from 5 GeV to 20 GeV. The second paper \cite{Drewes:2019fou} studies long-lived sterile neutrino in displaced
vertex searches and also using muon chambers to detect the muons from such vertex. 
It uses invariant mass cut and displacement cut to suppress the SM background. The third paper \cite{Bondarenko:2019tss} requires displaced vertex reconstructed by two muons in the CMS muon detector.

\textit{ \textbf{Acknowledgements:}}
We thank Felix Kling, Jared Evans and Bingxuan Liu for helpful discussion.
JL acknowledges support by an Oehme Fellowship. ZL is supported in part by the NSF under Grant No. PHY1620074 and by the
Maryland Center for Fundamental Physics. XPW is supported by the U.S. Department of Energy under Contract No. DE-AC02-06CH11357. LTW is supported by the DOE grant DE-SC0013642.
ZL thanks the Aspen Center for Physics, which is supported by National Science Foundation grant PHY-1607611.

\bibliographystyle{utphys}
\bibliography{reference}

\end{document}